\documentclass{aa}  
\usepackage{graphicx}
\usepackage{txfonts}
\begin{document}

\title{Dissecting the RELICS cluster SPT-CLJ0615-5746 through the intracluster light: confirmation of the multiple merging state of the cluster formation.}

\titlerunning{ICL and X-ray analysis of SPT0615}
\authorrunning{Jim\'enez-Teja et al.}

\author{Y. Jim\'enez-Teja \inst{1,2}\fnmsep\thanks{yojite@iaa.es} \and R. A. Dupke \inst{3,2,4} \and P. A. A. Lopes \inst{5} \and J. M. V\'ilchez \inst{1}}
\institute{Instituto de Astrof\'isica de Andaluc\'ia--CSIC, Glorieta de la Astronom\'ia s/n, E--18008 Granada, Spain\label{inst1} 
\and Observat\'orio Nacional, Rua General Jos\'e Cristino, 77 - Bairro Imperial de S\~ao Crist\'ov\~ao, Rio de Janeiro, 20921-400, Brazil\label{inst2}
\and Eureka Scientific Inc., 2452 Delmer Street Suite 100
Oakland, CA 94602-3017, USA\label{inst3}
\and Department of Astronomy, University of Michigan, 311 West Hall, 1085 South University Ave., Ann Arbor, MI 48109-1107\label{inst4}
\and Observat\'orio do Valongo, Universidade Federal do Rio de Janeiro, Ladeira do Pedro Ant\^onio 43, Rio de Janeiro RJ 20080-090, Brazil\label{inst5}}

\abstract{
The intracluster light (ICL) fraction, measured at certain specific wavelengths, has been shown to provide a good marker for determining the dynamical stage of galaxy clusters, i.e., merging versus relaxed, for small to intermediate redshifts. Here, we apply it for the first time to a high-redshift system, SPT-CLJ0615-5746 at $z=0.97$, using its RELICS  {(Reionization Lensing Cluster Survey)} observations in the optical and infrared.  We find the ICL fraction signature of merging, with values ranging from 16 to 37\%.  {A careful re-analysis of the X-ray data available for this cluster points to the presence of at least one current merger, and plausibly a second merger.} These two results are in contradiction with previous works based on X-ray data, which claimed the relaxed state of SPT-CLJ0615-5746, and confirmed the evidences presented by kinematic analyses. We also found an abnormally high ICL fraction in the rest-frame near ultraviolet wavelengths, which may be attributed to the combination of several phenomena such as an ICL injection during recent mergers of stars with average early-type spectra, the reversed star formation-density relation found at this high redshift in comparison with lower-redshift clusters, and projection effects.}


\maketitle
\nolinenumbers
\section{\bf Introduction}\label{sect_intro}

According to the standard cosmological paradigm Lambda-Cold Dark Matter \citep[$\Lambda$CDM, ][]{peebles2003}, clusters of galaxies are the latest and most massive structures to form in the Universe. Their process of formation and evolution follows a hierarchical structure, through the accretion of matter in filaments of the cosmic web, the addition of nearby galaxies, and the merging with groups and other clusters. Thus, their cluster formation pathway is plagued with interactions between galaxies and also with the intracluster medium. As a direct consequence of these violent processes, part of the stars in the interacting galaxies is freed to the intracluster space, just bound by the gravitational potential of the cluster, composing the so-called intracluster light (ICL). The ICL appears as an extended diffuse light in between 
the galaxy members of the clusters and especially concentrated around the brightest cluster galaxy (BCG), with a typical low surface brightness of $\mu_V>26.5$ mag arcsec$^{-2}$ \cite[e.g., ][]{rudick2006}. Thus, the ICL properties, from morphology to stellar composition, are intimately linked to the characteristics of its progenitor galaxies and can provide us with valuable information about the dynamics that govern the system as a whole. For example, it is estimated that during major galaxy mergers with the BCG, which generally happen at $z>1$, approximately between 30\% to 50 \% of the stars of the merging body become unbound and end up in the ICL \citep{murante2007, conroy2007, lidman2012,joo2023}. As a consequence,  {the color and metalicity radial profiles of the ICL are flat} (or with a very shallow gradient) in color and metallicity, reflecting the mixture of stellar populations that have been thrown to the intracluster space. On the other hand, at intermediate and low redshifts ($z<0.5$) the main mechanisms of ICL formation are believed to be cluster-cluster mergers, tidal stripping of luminous galaxies,  shredding of dwarf galaxies, and preprocessing in infalling groups \cite[e.g.][]{demaio2015, melnick2012, rudick2006,rudick2011, puchwein2010, murante2007, conroy2007,montes2014,montes2018,morishita2017,jimenez-teja2018,jimenez-teja2019,jimenez-teja2021,deoliveira2022,ragusa2023}. These processes originate color gradients in the ICL: 1) massive galaxies loose first the bluer stars located in their outer layers and, progressively as they orbit towards the potential well of the cluster or continue interacting with another galaxies, they free the redder stars of their inner regions;  {2) patches of ICL appear in those regions where groups are entering the gravitational potential of the cluster \citep[e.g.][]{iodice2017}; and 3) gravitational forces can act on dwarf galaxies at different clustercentric radii, originating a bluer ICL in the outskirts of the cluster as lower-mass dwarfs suffer the gravitational effect of the potential of the cluster at larger radii than more massive, more metal-rich dwarfs \citep{demaio2018}. Indeed, there is observational evidence indicating that the projected number density of dwarf galaxies anticorrelates with clustercentric distance, which may be the consequence of this transfer of material to the ICL \citep[e.g.,][]{venhola2018}. Moreover, observations of nearby clusters confirm that red low surface brightness dwarfs are more concentrated in the center of clusters, show more indications of disturbance, and tend to have smaller apparent axis ratios than normal dwarfs, which is also consistent with the cluster tidal forces acting on them and relocating their stars to the ICL \citep{lim2020,venhola2022}}. \\

Although different mechanisms have been identified as primary at different redshifts, only a few works in the literature have indeed analyzed the ICL of clusters at high redshift. It is believed that there is burst in the ICL formation at $z<0.5$ \citep{montes2018}, but abundant ICL is already set up at $z>1$ \citep[e.g.][]{burke2012,ko2018,joo2023} and even at $z>2$ \citep{adami2013}. In fact, different mechanisms are claimed to be primary depending on the redshift, being the ICL formation mostly linked to BCG build-up at $z> 1$ and to other channels related with satellite galaxies and infalling groups/galaxies at $z<1$. Interestingly, this two-phase scenario of formation implies that the ICL properties depend strongly on the instantaneous dynamical state of the cluster \citep{jimenez-teja2018}. This opened a new window to infer  the merging stage of clusters and even the age of particular systems called fossil groups \citep{dupke2022} as long as multiwavelength observations are available.\\

Here, we perform a multiwavelength study in optical  {and IR of the ICL fraction and in X-rays of the intracluster gas of} a high-redshift cluster observed by the Reionization Lensing Cluster Survey \citep[RELICS, ][]{coe2019}: SPT-CLJ0615-5746 at $z=0.972$ (SPT0615 hereafter),  {which has been one of the very few high-z clusters controversially classified as relaxed by some previous works \citep{bartalucci2019,planck2011,connor2019}.}\\


This paper is organized as follows. We firstly describe the optical and infrared data available for this clusters, its process of reduction, and algorithms to derive the ICL maps and fractions in Sect. \ref{sect_CICLEdata}. Next, we carry out an extensive analysis of available X-ray data of SPT0615 in Sec. \ref{sect:X-rays}. We discuss the results of the combined optical, infrared, and X-ray analysis in Sect. \ref{sect:discussion} and draw the main conclusions in Sect. \ref{sect:conclusions}. Throughout this paper we will assume a standard $\Lambda$CDM cosmology with $H_0=70$ km s$^{-1}$ Mpc$^{-1}$, $\Omega_m=0.3$, and $\Omega_{\Lambda}=0.7$. All magnitudes are referred to the AB system.\\

\section{Optical and infrared data and analysis}\label{sect_CICLEdata}

SPT0615 (R.A. = 6$^{\rm h}$15$^{\rm m}$56$^{\rm s}$.27, Dec. = -57$^{\circ}$45'50'' [J2000.0]) can be also found in the literature as PSZ1 G266.56-27.31. It is one of the most distant and massive clusters detected by Planck, at $z=0.972$ with  $M_{500}\sim 6.77^{+0.49}_{-0.54}\times 10^{14}\,M_{\odot}$ as estimated by this collaboration
\citep{ade2016}. SPT0615 was independently discovered by both the South Pole Telescope survey \citep{williamson2011} and the Planck Collaboration \citep{planck2011} using the Sunyaev–Zel’dovich (SZ) effect. It is exceptionally luminous, with a [0.1-2.4] keV band luminosity of $(22.7\pm0.8)\times 10^{44}$ erg s$^{-1}$, and hot, with $T_X\sim 11$ keV, \citep{planck2011}. \\

 We calculate intracluster light fractions (ICL fractions, defined as the ratio between the ICL and the total light of the cluster in a certain filter) for SPT0615 in six optical and infrared bands. To do so, we need to build an ICL map and an image that just contains the members of the cluster and the ICL for each filter considered. We used data gathered by the Hubble Space Telescope (HST) and the algorithm called CICLE \citep[CHEFs Intracluster Light Estimator, ][]{jimenez-teja2016}, which has been successfully applied before to high-quality HST data \citep{jimenez-teja2018,jimenez-teja2021,deoliveira2022,dupke2022} to build the ICL maps, and a machine learning algorithm \citep{lop20} to perform the cluster membership.\\ 

\subsection{HST data}\label{sect_HSTdata}

RELICS \citep{coe2019} is a multi-orbit Hubble Treasure Program devoted to observe the 21 most massive and distant clusters of galaxies, according to Planck estimations \citep{planck2016}, as well as 20 additional systems selected by their strong lensing nature. RELICS provides information both in the optical with the Advanced Camera for Surveys (ACS) and in the infrared (IR) with the Wide Field Camera 3 (WFC3). Previous to the RELICS survey, there existed archival HST images of SPT0615 in the optical filters F606W and F814W (programs \#12477, PI: High, and \#12757, PI: Mazzotta). RELICS completed these observations with one HST orbit in the F435W filter and two orbits split among the four IR filters F105W, F125W, F140W, and F160W.\\

All images were reduced by the RELICS collaboration applying the standard pipelines CALACS\footnote{http://www.stsci.edu/hst/acs/performance/calacs\_cte/calacs\_cte.html} for the optical data and CALWF3\footnote{http://www.stsci.edu/hst/wfc3/pipeline/wfc3\_pipeline} for the IR images. These pipelines include corrections by bias, dark, flat-fielding, bias-striping, crosstalk, and charge transfer efficiency. Also, persistence models were calculated to mask those pixels in the IR images that could have some afterglow remaining from previous pointings. Although it is not optimal for low-surface-brightness studies, in \cite{jimenez-teja2021} we showed that the standard flat-field correction provided by the standard HST pipelines does not affect the resulting ICL fractions if total fluxes are measured. However, for those images where we observed that the standard MAST flat-field image could be compromising the ICL measurement, we calculated new sky-flats using the prescriptions described by \cite{borlaff2019}. Individual exposures where later aligned and combined using Astrodrizzle \citep{koekemoer2002} to a pixel scale of 0.06 arcsec. We estimated the limiting surface brightness in 3$\times$3 arcsec$^2$ boxes for each filter using the prescription described by \citet{roman2020} (see Table \ref{table:SBlimits_ICLfractions}).\\

\begin{table*}
\centering
\begin{tabular}{cccccc}
Filter & Surface brightness limit & ICL fraction & & Errors &\\
& & & Photometric & Cluster membership & Geometrical \\
   & [mag arcsec$^{-2}$] & [\%] & [\%] & [\%] & [\%] \\
   \hline
F606W &  {26.73 $\pm$ 0.06} & 36.3 $\pm$ 5.0 &  {2.5} &  {0.9} &  {1.6}\\
F814W &  {26.44 $\pm$ 0.09} & 26.6 $\pm$ 1.6 &  {0.9} &  {0.6} &  {0.2}\\
F105W &  {26.02 $\pm$ 0.28} & 23.3 $\pm$ 1.7 &  {0.5} &  {0.5} &  {0.8}\\ 
F125W &  {25.87 $\pm$ 0.42} & 22.5 $\pm$ 0.9 &  {0.5} &  {0.4} &  {0.0}\\
F140W &  {25.92 $\pm$ 0.31} & 16.6 $\pm$ 3.2 &  {0.3} &  {0.3} &  {2.5}\\
F160W &  {25.44 $\pm$ 0.23} & 21.0 $\pm$ 1.2 &  {0.4} &  {0.4} &  {0.4}\\
\hline
\end{tabular}
\caption{ {Limiting surface brightness calculated in boxes of 3$\times$3 arsec$^2$, ICL fractions computed with CICLE, and breakdown of the ICL fraction error into the three sources considered}.} \label{table:SBlimits_ICLfractions}
\end{table*}

\subsection{Generation of the ICL maps}\label{sect_ICLmaps}

In a nutshell, CICLE \citep{jimenez-teja2016} eliminates the galactic luminous contribution by masking the stars and fitting the galaxies. Stars are masked either using SExtractor segmentation map \citep{bertin1996} or manually, when the pixels assigned by SExtractor to the star clearly not cover their full extension. Galaxies are modeled using a combination of Chebyshev rational functions and Fourier series that form a mathematical orthonormal basis capable of fitting objects with any kind of morphology, given that they are smooth enough \citep{jimenez-teja2012}. For instance, saturated stars or objects 
located too close to the border images are excluded from the set of morphologies that CHEFs can fit. The CHEF algorithm determines
the extension of  {stellar halo} of the galaxies modeled by estimating, on each radial direction, the point where the projected galactic surface either submerges into the sky or converges asymptotically to a certain value.  {The ICL tends to follows the gravitational potential of the cluster, so it is usually more concentrated around its center, close to the BCG. As a consequence, this procedure to find the observational limits of the stellar haloes is only valid for galaxies located relatively far from the cluster center. Thus, it cannot be applied to disentangle the BCG from the ICL, as their projected distributions are usually aligned. For this particular case, CICLE uses a curvature parameter, called minimum principal curvature \citep{patrikalakis2010}, to find the limits of the region dominated by the BCG over the ICL. Intuitively, this parameter quantifies the change in the slope of a surface, point by point. CICLE calculates the curvature map of the composite BCG+ICL surface and identifies the maxima values (i.e., the points where it finds the highest change in the slope) with the transition from the BCG- to the ICL-dominated area. Obviously, the more dissimilar the profiles or the steepness of the BCG and the ICL are, the easier to find this transition. One must note, though, that this does not mean that the whole stellar halo of the BCG is contained within this region, but a small amount of it will still extend over the ICL-dominated region, ``covered'' by the ICL in projection. The BCG flux missed in this way is included in the final error budget, as described in Sect. \ref{sect_ICLfanderror}. After modelling the BCG within this region, we obtain a map that just contains ICL and background}.\\

\begin{figure*}
\centering
\includegraphics[width=.48\textwidth]{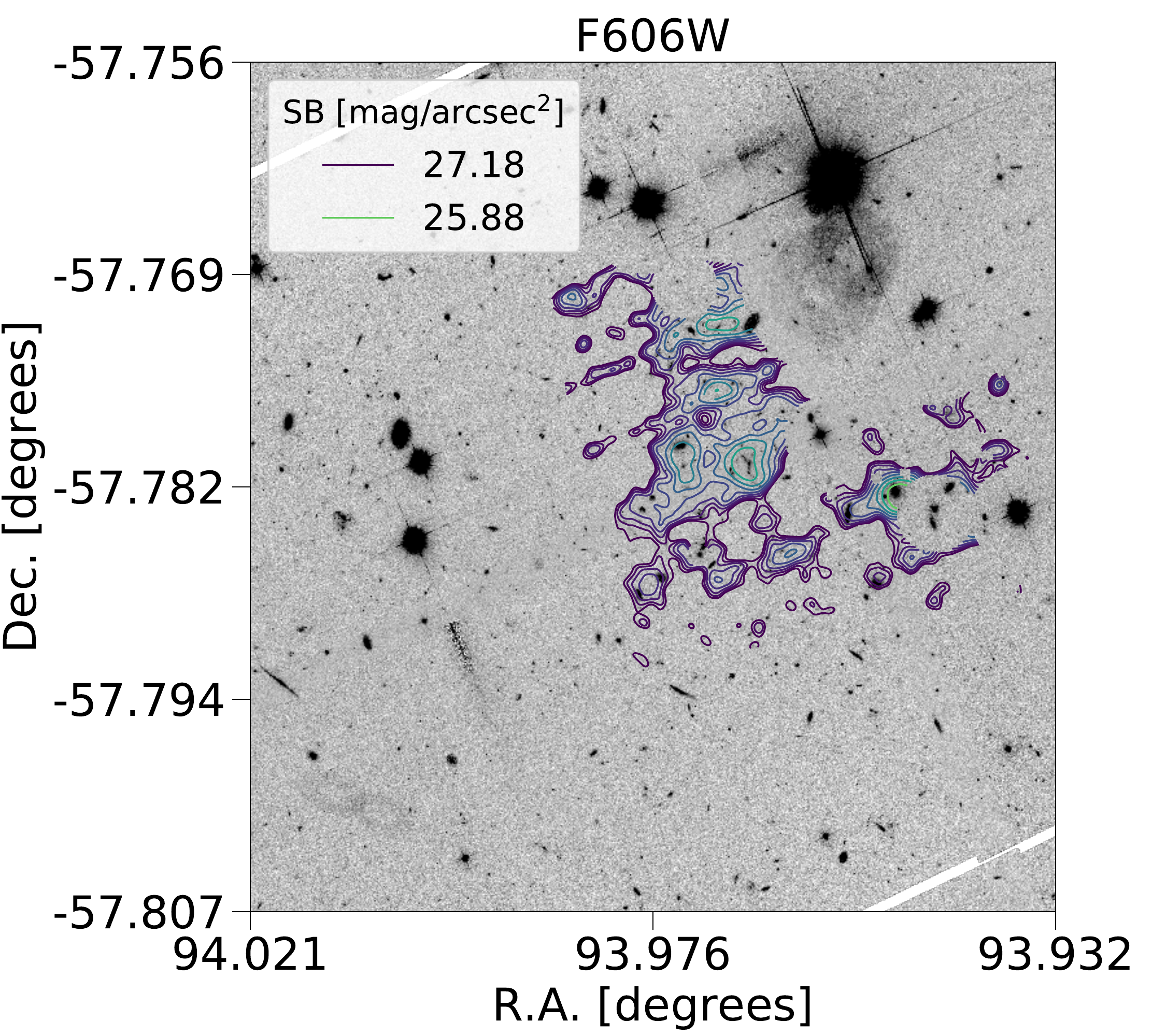}\hfill\includegraphics[width=.48\textwidth]{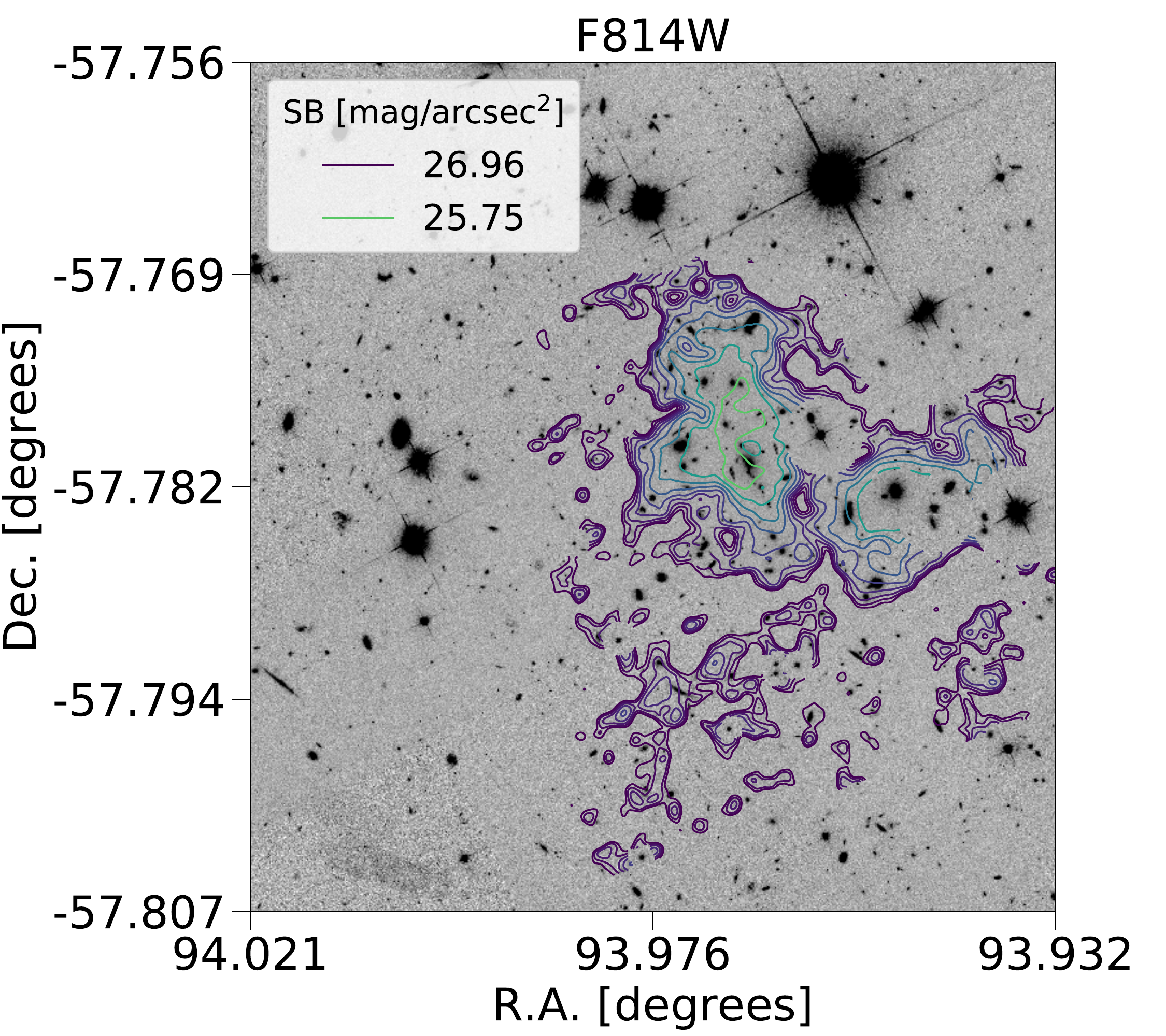}\hfill \\
\includegraphics[width=.48\textwidth]{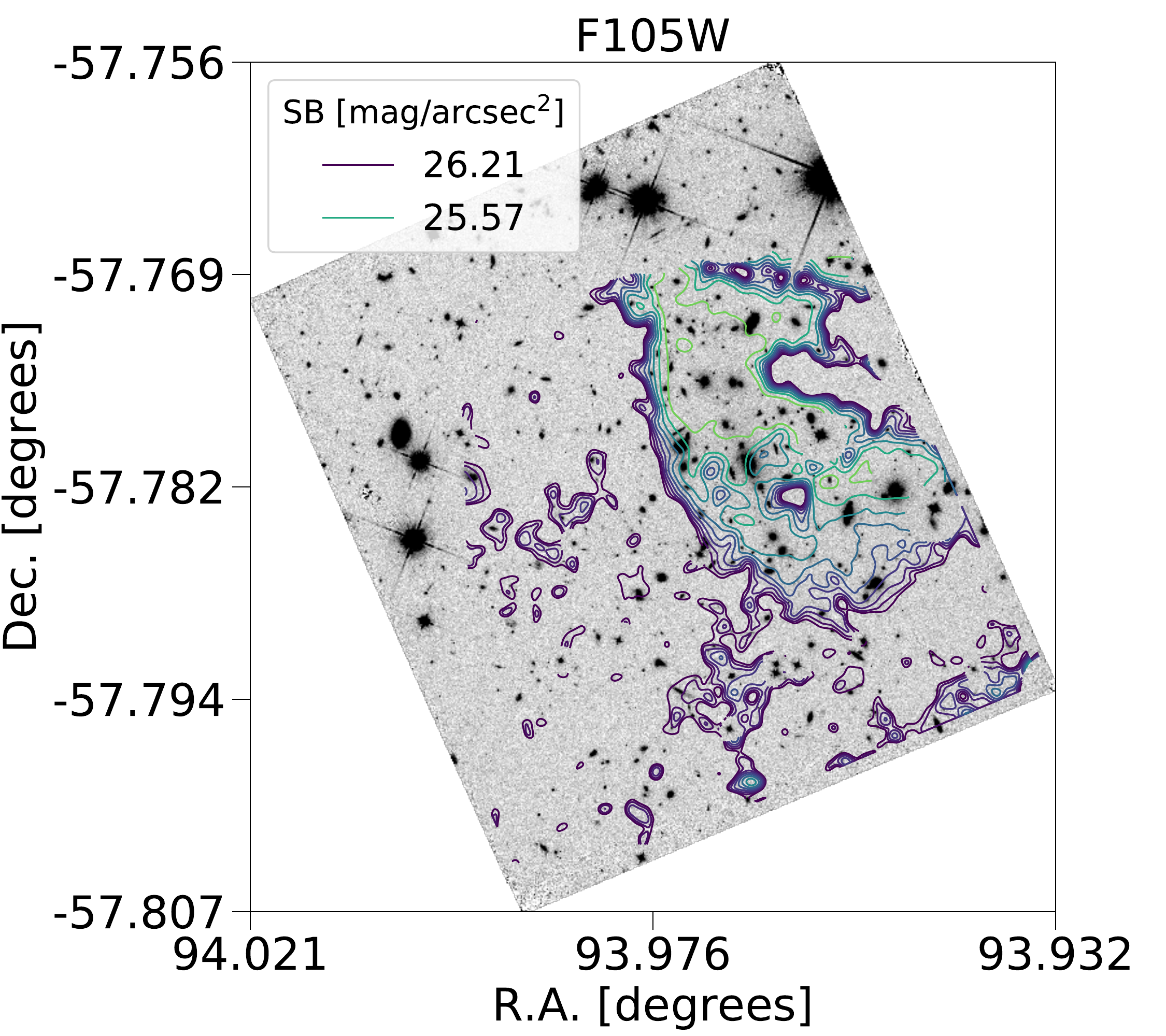}\hfill\includegraphics[width=.48\textwidth]{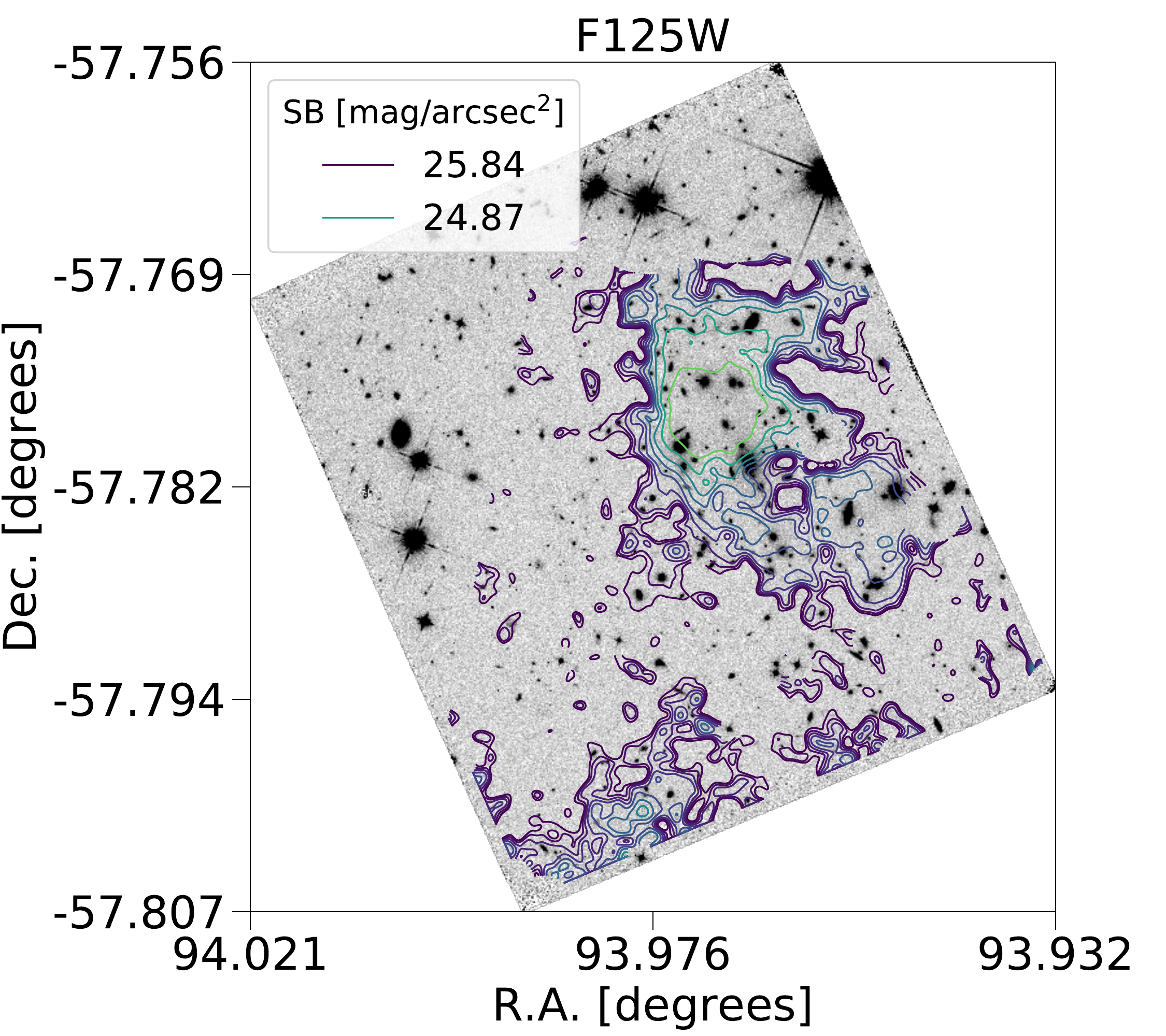}\hfill\\
\includegraphics[width=.48\textwidth]{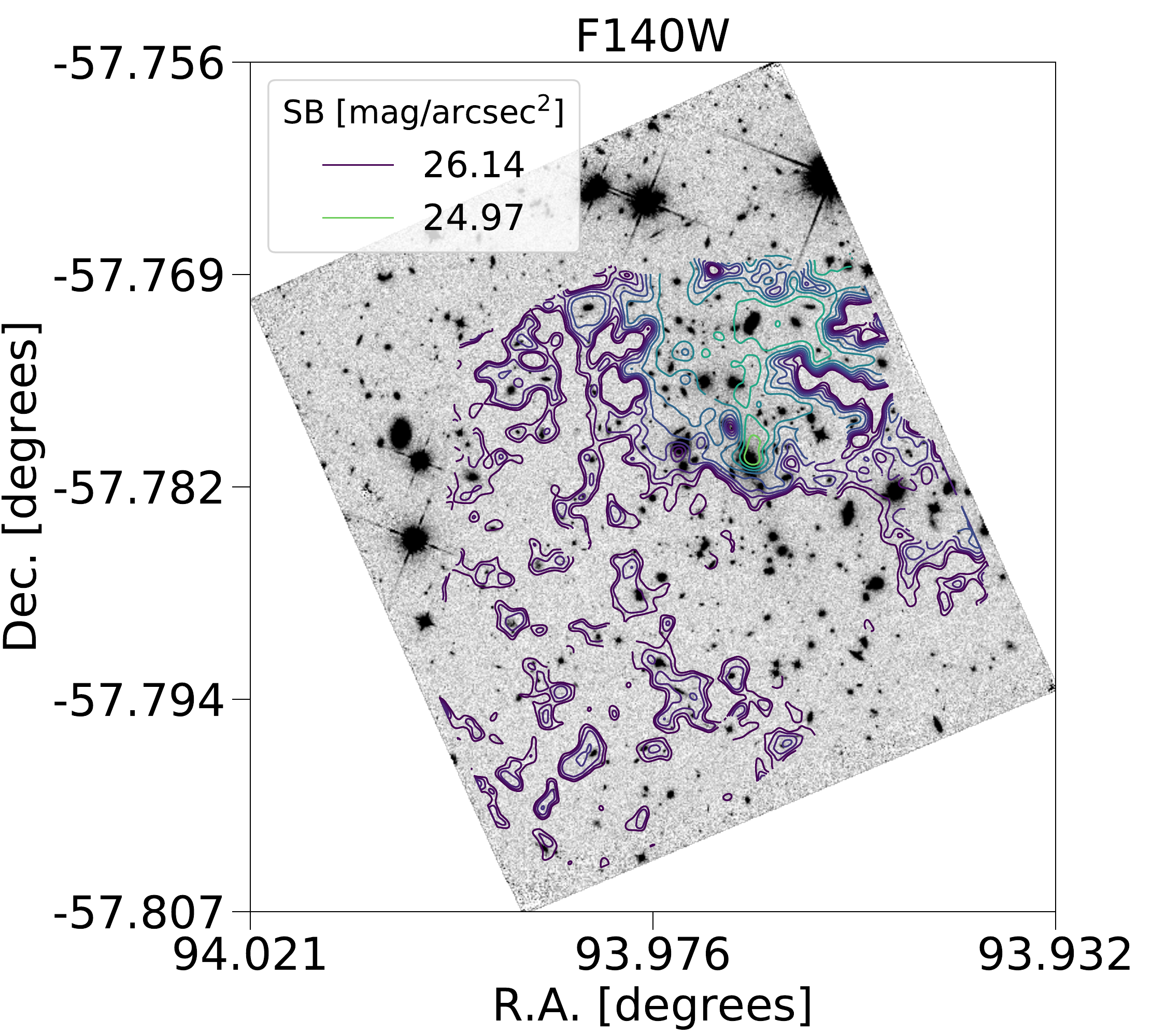}\hfill\includegraphics[width=.48\textwidth]{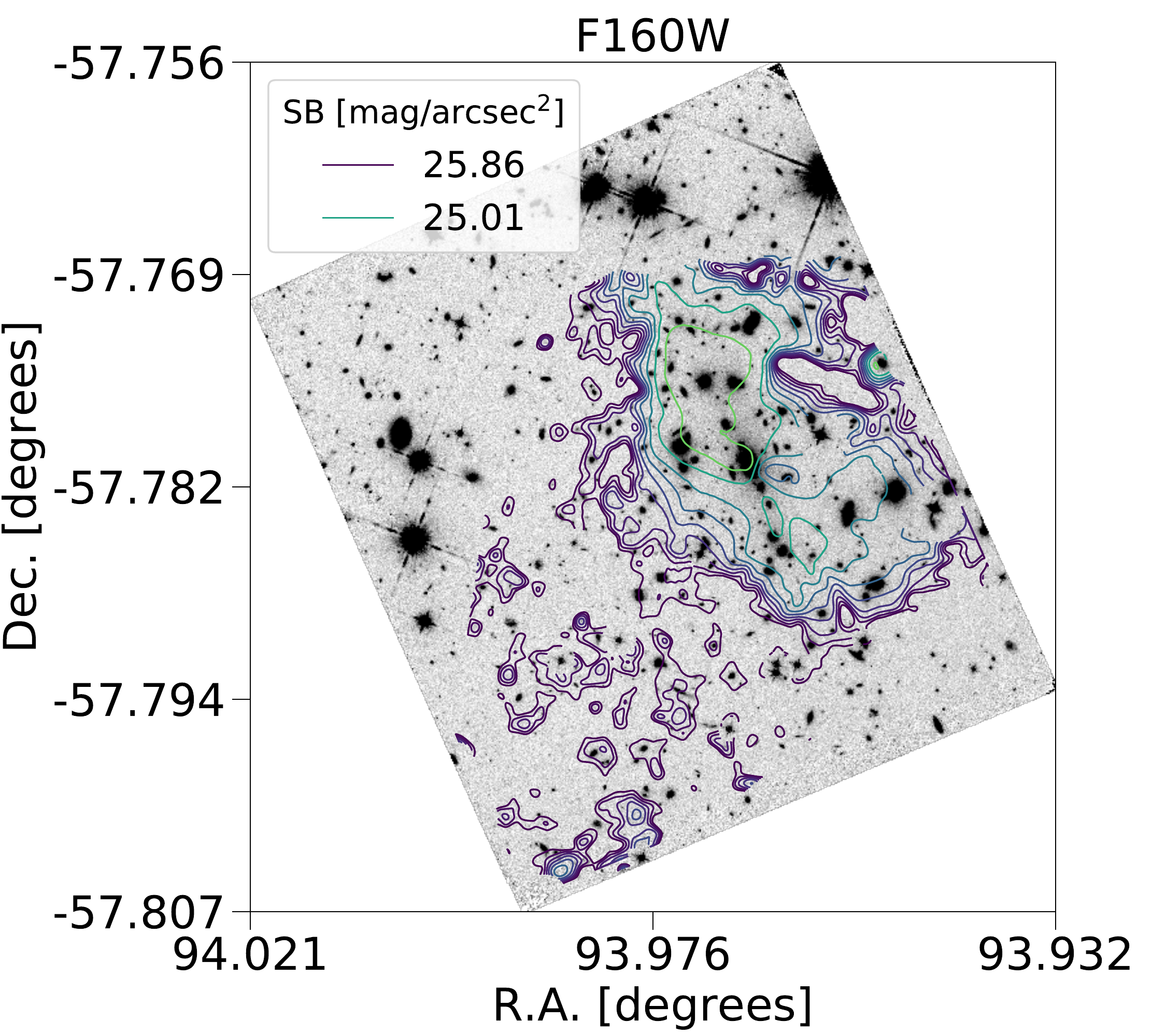}\hfill
\caption{ICL isocontours superimposed over the original images in the six HST ACS and WFC3 filters considered for this work. For each band, we plot 10 isocontours logarithmically spaced, where the lowest level is calculated from to the detection limit of the ICL (where it converges to the background level) and the highest corresponds to its maximum value. Two main clumps of ICL appear in all bands, more clearly separated in the bluer filters.}
\label{ICLcontours}
\end{figure*}

Although calibrated images already have the sky removed, it is mandatory to perform a refinement of this background in a final step. We can find in the literature many different approaches for this estimation of the background, such as using nearby fields \citep[observed under the same observational and technical characteristics, e.g. ][]{jimenez-teja2016}, analyzing the distribution of the pixels that are considered as ICL-free \citep{morishita2017,jimenez-teja2021}, or using specialized softwares as SExtractor \citep[e.g., ][]{burke2012}, or, more recently, NoiseChisel \citep{borlaff2019}. To our understanding, the first one would be the safest approach, but we do not usually have parallel fields with the same observational characteristics available. Studying the distribution of the blank pixels in the image is not feasible in our case due to the local variations in the background (in spite of having recalculated the sky flats, as decsribed in Sect. \ref{sect_HSTdata}) and the larger field of view associated to some of the filters considered in this work. \cite{borlaff2019} made an exhaustive study of the different ways to estimate the sky and concluded that, although all algorithms tend to overestimate the sky level, NoiseChisel \citep{akhlaghi2015} was the one that performed the best. More recent works with different datasets have led to similar conclusions \citep{haigh2021,kelvin2023}. We used the updated version of NoiseChisel \citep{akhlaghi2019} to calculate our background maps. Then, we remove them from the ICL+background images previously calculated to obtain the final ICL maps. In Fig. \ref{ICLcontours}, we show the contours of the ICL in the six optical and IR filters analyzed in this work, superimposed over the original images. We observe an overall similar morphology of the ICL in the different bands, elongated along a north-south axis with a position angle of 30º approximately. It also appears to be preferentially concentrated around two main clumps, north and south-west to the BCG respectively.\\

\subsection{A Machine Learning Approach to Photometric Membership}\label{sect_clustermembership}

The second step is to generate an image with the stellar content of the cluster, that is, the ICL and the cluster galaxies. We just add the CHEF models of the galaxies to the ICL map, after determining which galaxies belong to the cluster. We applied the code named Reliable Photometric Membership
(RPM, \citealt{lop20}) to select galaxy cluster members based only in
photometric parameters of the galaxies lying along the line of sight of
clusters. RPM employs a machine learning (ML) approach to derive membership
probabilities, which are then used to derive a membership classification.
In \citet{lop20} the code efficiency was verified for low redshift systems,
within $R_{200}$. In \citet{jimenez-teja2021} we demonstrated the code also works for
higher-z systems ($z < 1$), within the RELICS survey footprint.\\

The ML method was trained and evaluated with galaxies within the regions of
18 CLASH clusters, covering a broad redshift range ($0.0792 < z < 0.8950$).
The best results we achieved considered galaxies within 1.50 h$^{-1}$ Mpc of
the cluster centre and with 15 $\le$ F814W $\le$ 25. Our final data set
(used to train and validate the ML method) comprises 927 galaxies with
spectroscopic redshifts, from the CLASH clusters. Further details on these
steps can be found in \citet{jimenez-teja2021}.\\

As discussed in \citet{jimenez-teja2021} we found that the Stochastic Gradient Boosting
(GBM) method results in a slightly better performance, when compared to other
ML models. That is quantified by the completeness (also called 
"True Positive Rate", TPR or "Sensitivity") and purity (known as "Precision"
or "Positive Predictive Value", PPV). These two parameters track the relation
between the sample of objects classified as members and the true population
of members.\\

The photometric parameters employed by the ML model are $\Delta$ (F435W-F814W), $\Delta$ (F606W-F814W), $\Delta$ (F105W-F140W), $\Delta$ (F814W-F125W), $\Delta$ (F814W-F140W), LOG $\Sigma_5$, and $\Delta_{z\,\text{phot}}$. $\Delta$ stands for
the offset relative to the mean magnitude, color or cluster redshift. We do not
use apparent magnitudes or observed colors due to the large redshift range
of the training sample. We also avoid absolute magnitudes and rest-frame
colors that may be subject to large uncertainties associated to the photo-z
precision and due to the $k-$ and $e-$corrections.\\

We achieve high values of Completeness (C) and Purity P,
C $ = 93.5\% \pm 2.4\%$ and P $ = 85.7\% \pm 3.1\%$. This methodology was
sucessfully applied to all galaxies in the regions of the 25 CLASH clusters,
as well as for 35 of the 42 RELICS clusters, including WHLJ013719.8-082841
(for which the ICL was investigated in \citet{jimenez-teja2021}). For the particular case of SPT0615, 174 galaxies are photometrically classified as cluster members. Additionally, we have spectroscopic redshifts for 58 galaxies  \citep{connor2019} and 47 out of them have a similar redshift as that of the cluster (having velocity offsets smaller than 5000 km/s). We then used a ``shifting gapper'' technique \citep{fad96, lop09} to select members and exclude interlopers in the projected
phase space.  {That is simply based on the application of the gap-technique in radial bins from the cluster center (described in \citealt{katgert96}), to identify gaps in the redshift (velocity) distribution. Instead of adopting a fixed gap, such as 1000 km s$^{-1}$, we considered a variable gap, called density gap \citep{adami98, lop09}, which depends on the number density of galaxies in the cluster region.} Thus, according to the spectroscopic information, we found that
39 galaxies are members of SPT-CLJ0615-5746, while 8 are interlopers. Regarding
the photometric membership, 32 out of those 39 are also selected as
members ($\sim 82\%$). That gives a lower limit to our completeness, as we
have many more galaxies with photometry than spectra in that region. \\

\begin{figure}
\centering
\includegraphics[width=.48\textwidth]{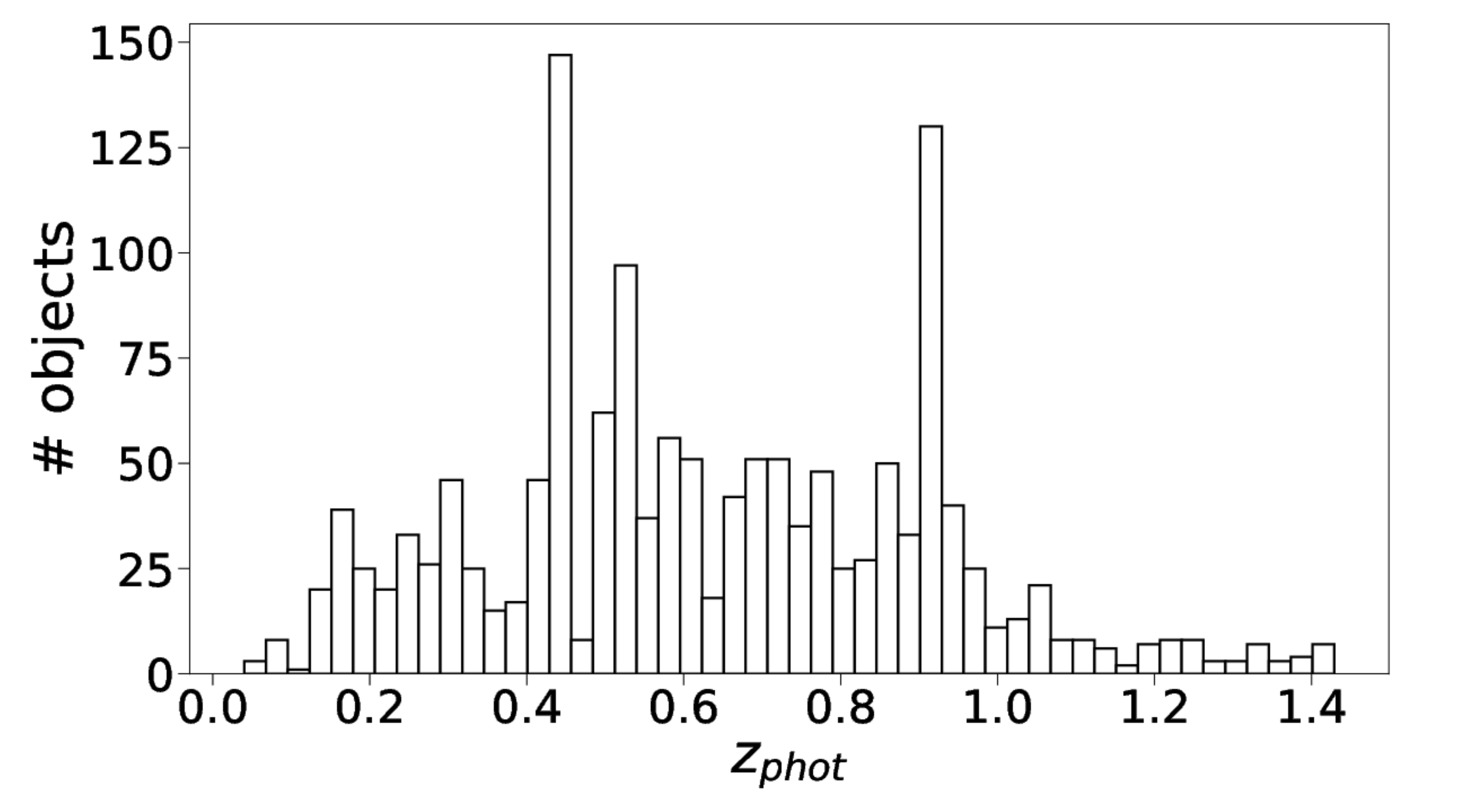}\\
\includegraphics[width=.48\textwidth]{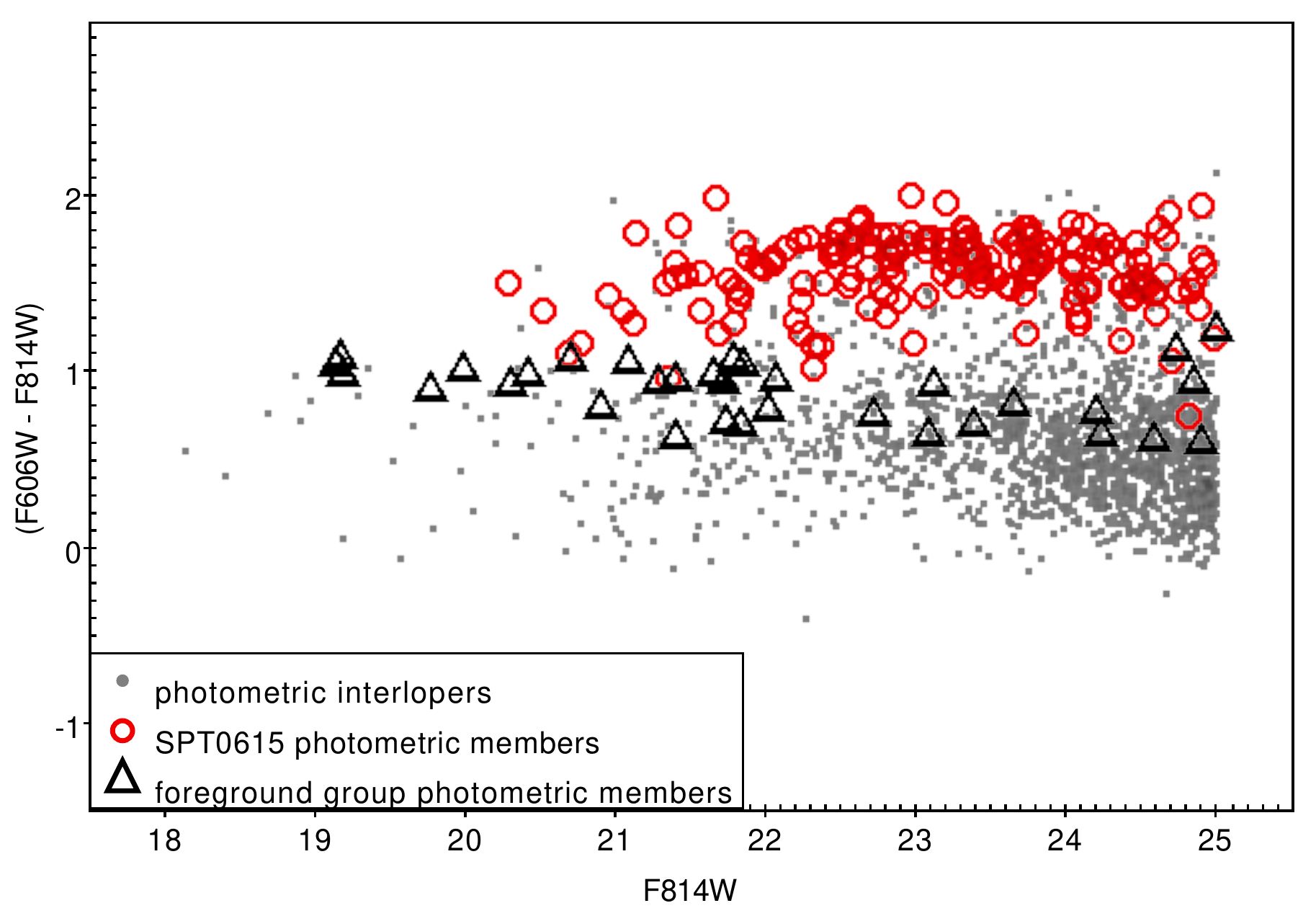}
\caption{Top: Photometric redshift distribution of the galaxies in the region of SPT0615. Bottom: Color-magnitud diagram with the members of SPT0615 at $z=0.97$ in red and those of the possible foreground structure at $z\sim 0.44$ in black. Gray points correspond to interlopers that do not belong to any of the two structures.}
\label{z_dist}
\end{figure}

One interesting point faced in the study of SPT0615 is the presence of a
possible foreground structure at $z \sim 0.4$, as pointed out by \citet{paterno-mahler2018}. To investigate this, we explored the photometric redshift distribution of the galaxies in the region of SPT0615 (Fig. \ref{z_dist} top) and we verified that is very likely that a second system is projected along the line of sight, at $z_{phot} \sim 0.44$.  Hence, we decided to apply the photometric membership selection also taking this
foreground structure into account. We repeated the procedure described above for all the galaxies in the footprint of SPT0615, but now considering the centroid and photometric redshift of this possible foreground group as reference. We calculated two different probabilities for each galaxy to either belong to SPT0615 or to the structure at $z_{phot} \sim 0.44$. For each object, we included  the highest membership probability of these two in our final catalog and classified it as  member or interloper relative to the high$-z$ cluster and to the foreground group. The color-magnitud diagram (CMD, see Fig. \ref{z_dist} bottom), shows a possible second red sequence with three brightest galaxies of very similar magnitude. Two of them are very close, southwest to SPT0615 BCG. We took the middle between these two galaxies as center of the possible second structure. If exists, the foreground cluster or group would have 38 members identified.  {We check the spatial distribution of this putative foreground group in Sect. \ref{sect:discussion}}.\\

\subsection{Calculation of the ICL fraction and error}\label{sect_ICLfanderror}

\begin{figure}
\centering
\includegraphics[width=.5\textwidth]{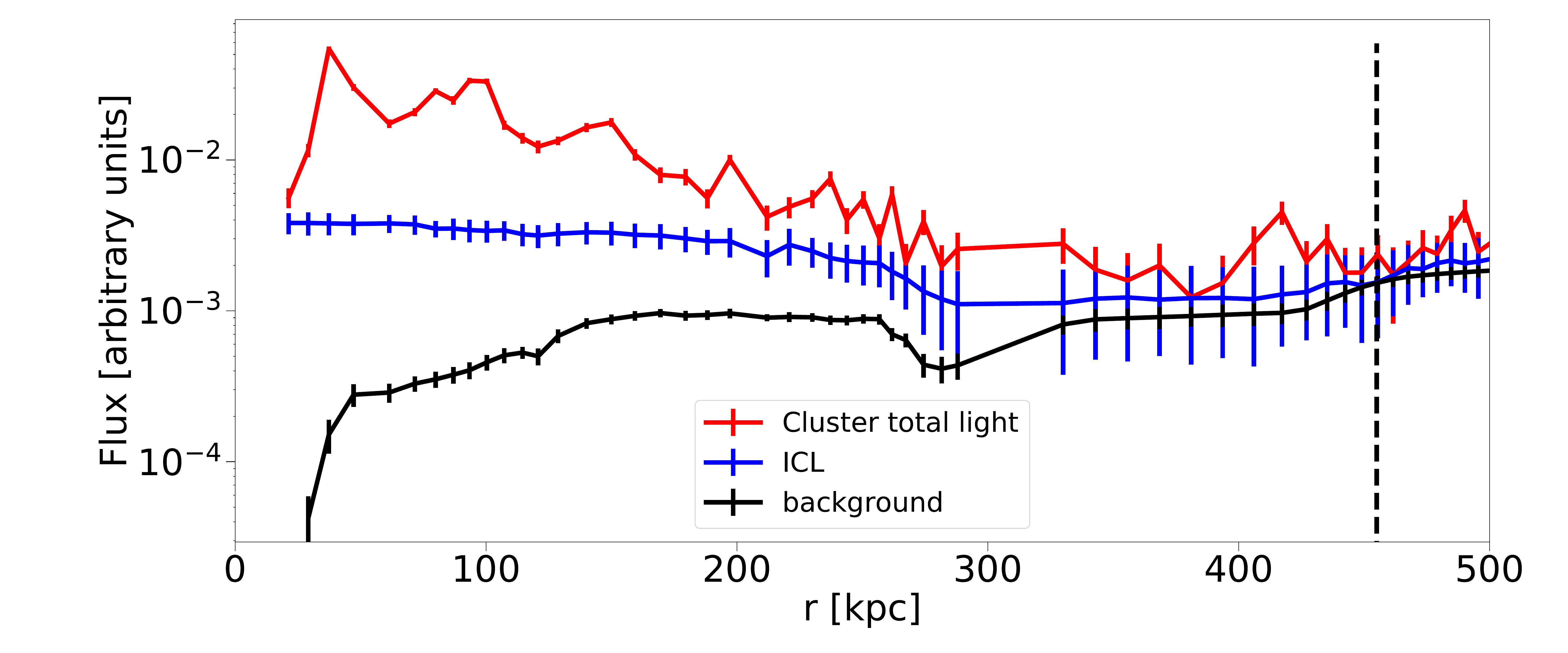}
\caption{ {Radial profiles of the total cluster, the ICL, and the background in the F160W filter. As the background was negative in some regions, we have added an arbitrary quantity to the three profiles to be able to plot the y axis in logarithmic scale. Dashed vertical line indicates the limit of the ICL.}}
\label{fig:F160W_profile}
\end{figure}

We define the limit of the ICL as those points where the ICL submerges into the sky level.  {We first calculate the radial profile of the ICL and the background and search for the average radius where they coincide. For the sake of illustration, we plot these radial profiles (along with that of the total cluster) for the F160W band in Fig. \ref{fig:F160W_profile}. Errors are drawn from jacknife resampling. We can observe how the radial profile of the ICL decreases until it reaches that of the background at a radius of $r=455$ kpc.} We then calculate our final ICL fractions measuring the total flux of the ICL and the total cluster light, within this region. We note here that this limit is likely non-physical, but it is highly dependant on the depth and the size of our observations, as proved by recent simulations \citep{deason2021}. As opposed to measuring the ICL in fixed apertures, total ICL fractions are independent of the point spread function (PSF) and have the advantage of diluting the possible systematic errors inherent to the calibration and reduction of the images given the larger areas involved in the calculations \citep{jimenez-teja2021}.  {The error in the ICL fraction is estimated considering three different contributions: the photometric, the geometrical, and cluster membership errors. The photometric error is that inherent to the process of measuring the flux of a source with noise, and depends on the flux measured, the gain, the rms of the sky, and the area covered by the source. We call geometrical error that made by CICLE in the disentanglement of the BCG from the ICL. It accounts for both the error in the determination of the transition from the BCG- to the ICL-dominated regions and the flux missed from the BCG outer stellar halo that extends over the ICL-dominated region (see Sect. \ref{sect_ICLmaps}). This error depends on how different the profiles of the BCG and the ICL are (that is, their magnitudes and effective radii) and it is estimated using simulated images that mimic the geometrical configuration of the BCG and the ICL. Finally, we compute the error associated to the photometric cluster membership, which depends on the completeness yielded by the machine learning algorithm. From the resulting completeness of 93.5\% (see Sect. \ref{sect_clustermembership}), we estimate the flux missed from the missing cluster galaxies in each filter and we propagate it to the ICL fraction. The final ICL fractions are high, ranging between 16.6 and 33.3\%, as listed in Table \ref{table:SBlimits_ICLfractions}. Final errors, which range between 0.9 to 5.0\%, are also reported, along with the breakdown of the total error budget into the three sources of error described above}.\\

\section{Dynamical stage of SPT0615} \label{sect:X-rays}

As there exists a clear correlation between the ICL fraction and the dynamical stage of the cluster \citep{jimenez-teja2018}, we analyzed the available X-ray data and compared our results with those of previous works \citep[e.g., ][]{bartalucci2017,yuan2020,bulbul2019,planck2011}.

\subsection{X-ray data and analysis} \label{chandra}

All archived Chandra data available for SPT0615 were used. There are 12 separate observations of SPT0615 (OBS ID: 14017, 14018, 14349, 14350, 14351, 14437, 15572, 15574, 15579, 15582, 15588 and 15589) taken with the Advanced CCD Imaging Spectrometer (ACIS-I), from Sept. 15th to Nov 24th 2012 aimed at RA 06:15:52.0, DEC -57:46:51.6 (PI: Mazzotta). We used CIAO 4.13 to run the standard recommended data processing steps, reprocessing all the data with the script \textit{chandra\_repro}, which creates new bad pixel files and new level 2 event files. The total effective (after cleaning) exposure is 241 ksec. We created a merged event file, reprojecting all observations to a common tangent point using the \textit{reproject\_obs} tool. This was used solely for dealing with point source removal and basic image analysis  {after refilling the extracted point sources using the tool \textit{dmfilth}}. For the spectral analysis we used the individual event files to create separate data products and analyzed them simultaneously, following \textit{CIAO} guidelines\footnote{ cxc.cfa.harvard.edu/ciao/ahelp/reproject\_obs.html}. \\

The tool \textit{specextract} was used to produce spectral files and responses for the regions analyzed. Spectra was grouped to have from 5 to 20 cnt/channel depending on the region configuration being analyzed. Given the small source angular extension, we used two local background regions, away from the cluster’s center within the same CCD. One about 1.4 Mpc and the other 1.9 Mpc from the clusters’ center at different orientations with respect to the N-S cluster X-ray elongation, to test, among other things, the importance of cluster contamination near R$_{200}$. Spectral fittings were carried out  in \textit{XSPEC 12.11.0m},  using an absorbed Collisional Ionization Equilibrium model \textit{tbabs*apec}, with the  redshift 0.972 \citep{connor2019} fixed at their nominal value of the BCG and the Hydrogen column density (nH) of 3.2$\times10^{20} cm^{-2}$ chosen from the HI4PI Map  \citep{HI4PI} through the HEASARC \textit{nH} tool\footnote{heasarc.gsfc.nasa.gov/cgi-bin/Tools/w3nh/w3nh.pl}. Abundances listed here are with respect to the photospheric value \citep{angr}.  {For the 2-dimensional image fitting and production of residual emission we used Sherpa v.1 \cite{sherpa1} in Ciao 4.1.4.} \\
 
\begin{figure*}
\centering
\includegraphics[width=\hsize]{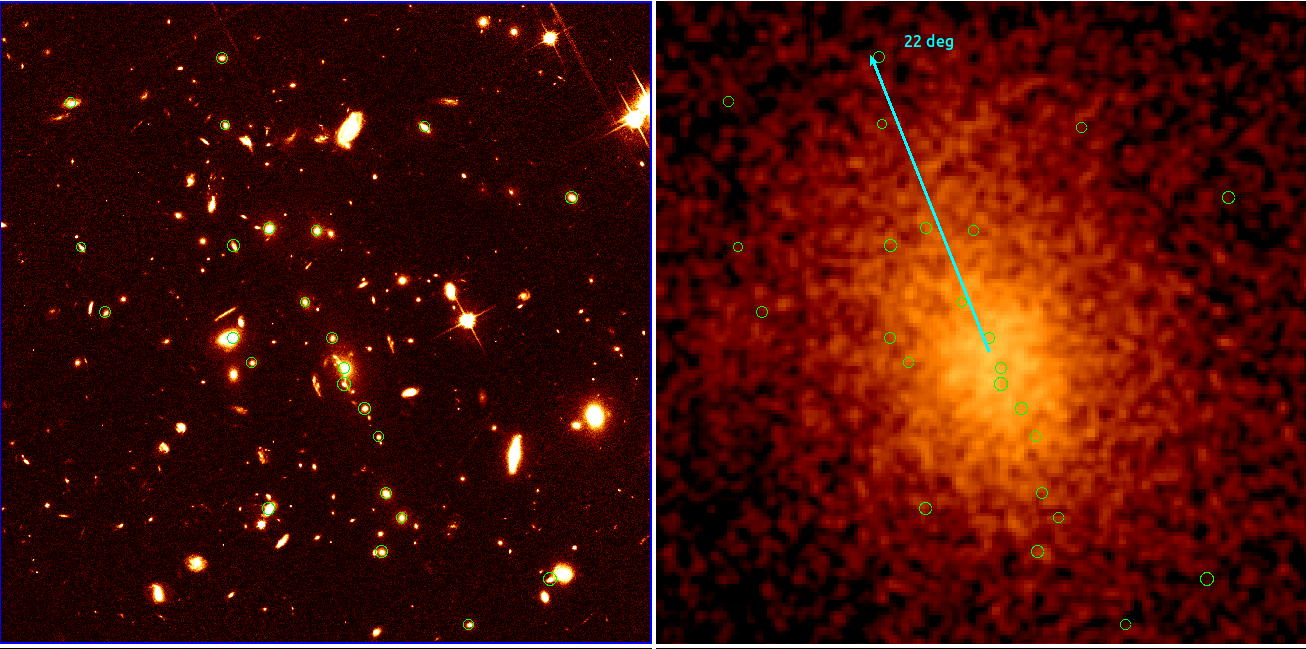}
\includegraphics[width=\hsize]{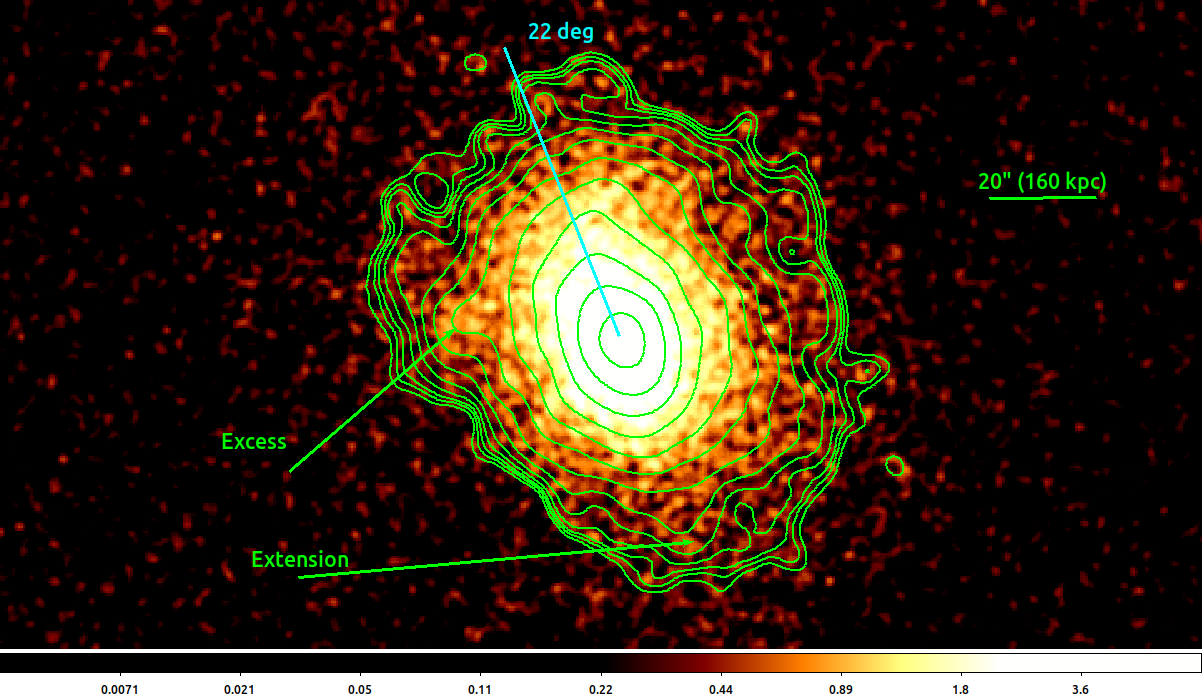}
\caption{Top left: HST/ACS image in the F814W filter of SPT0615. Spectroscopically confirmed member galaxies are pointed by green circles. Top right: Chandra image of the same region. The alignment seen in the galaxy distribution coincides with the major axis of the surface brightness distribution in X-rays, at an inclination of $\sim22^o$ shown in the bottom figure. Physical scale of 160 kpc. Bottom:  X-ray isocontours of SPT0615, which show a similar elongation as the galaxy distribution and lensing critical curves \citep[Figure 6 of][]{paterno-mahler2018}. Excess X-ray emission can be seen clearly in the East and South and are marked by green arrows. North is up. East is left.
}
\label{x-cont}
\end{figure*}

\subsection{X-ray results and temperature distribution}\label{sect_Xrayresults}

\begin{figure}
\centering
\includegraphics[width=9cm]{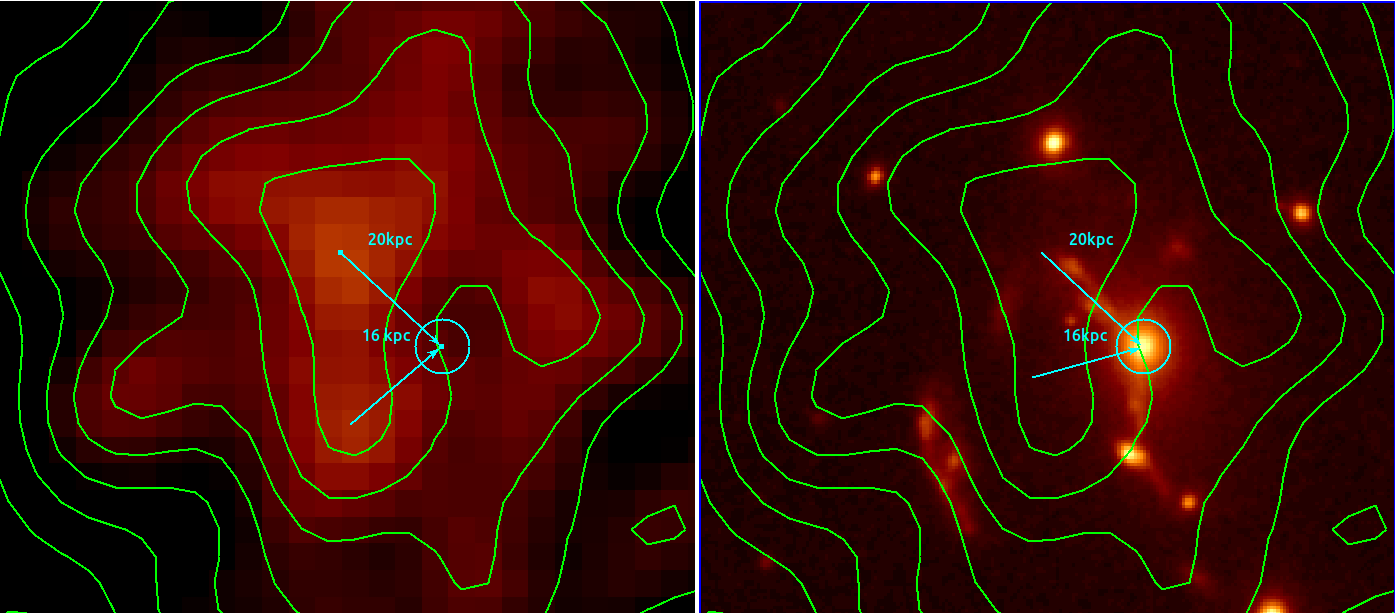}

\caption{Left: Chandra X-ray image and isocontours in the very core of SPT0615. The position of the BCG is shown by a circle as well as the approximate distance from each of the bright X-ray "cores".  Right:  HST image of the same region with X-ray contours overlaid. None of the X-ray brightest central regions coincide with the BCG or any other galaxy.
}
\label{core}
\end{figure}

Chandra image of SPT0615 shows a clear X-ray elongation along an North-South direction, where the main axis has a position angle of $\sim~22^{\circ}$ (Fig. \ref{x-cont} top right). Similar elongation is also seen in the   gravitational lensing critical curves 
\citep{paterno-mahler2018} and in the ICL distribution (Fig. \ref{ICLcontours}). Galaxy members in the cluster tend to be distributed along this direction as well (Fig. \ref{x-cont} top left).  {The X-ray surface brightness contours suggest also some asymmetric excess emission to the East and possibly Southern regions (Fig. \ref{x-cont} bottom; Sect. \ref{ximage}).}  At small scales, near the center, there are apparently two X-ray brightest peaks 
configured along the N-S direction, displaced from the BCG by 20 kpc and 16 kpc (Fig. \ref{core}).   \\

\begin{figure}\centering
\resizebox{7cm}{!} {\includegraphics{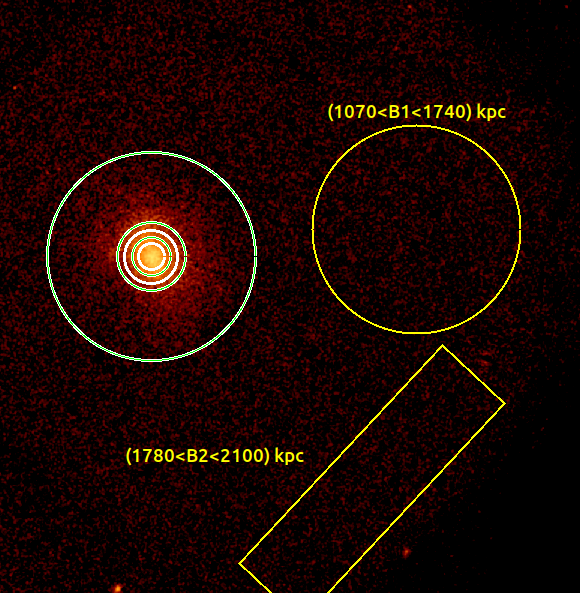}}
\resizebox{7cm}{!} {\includegraphics{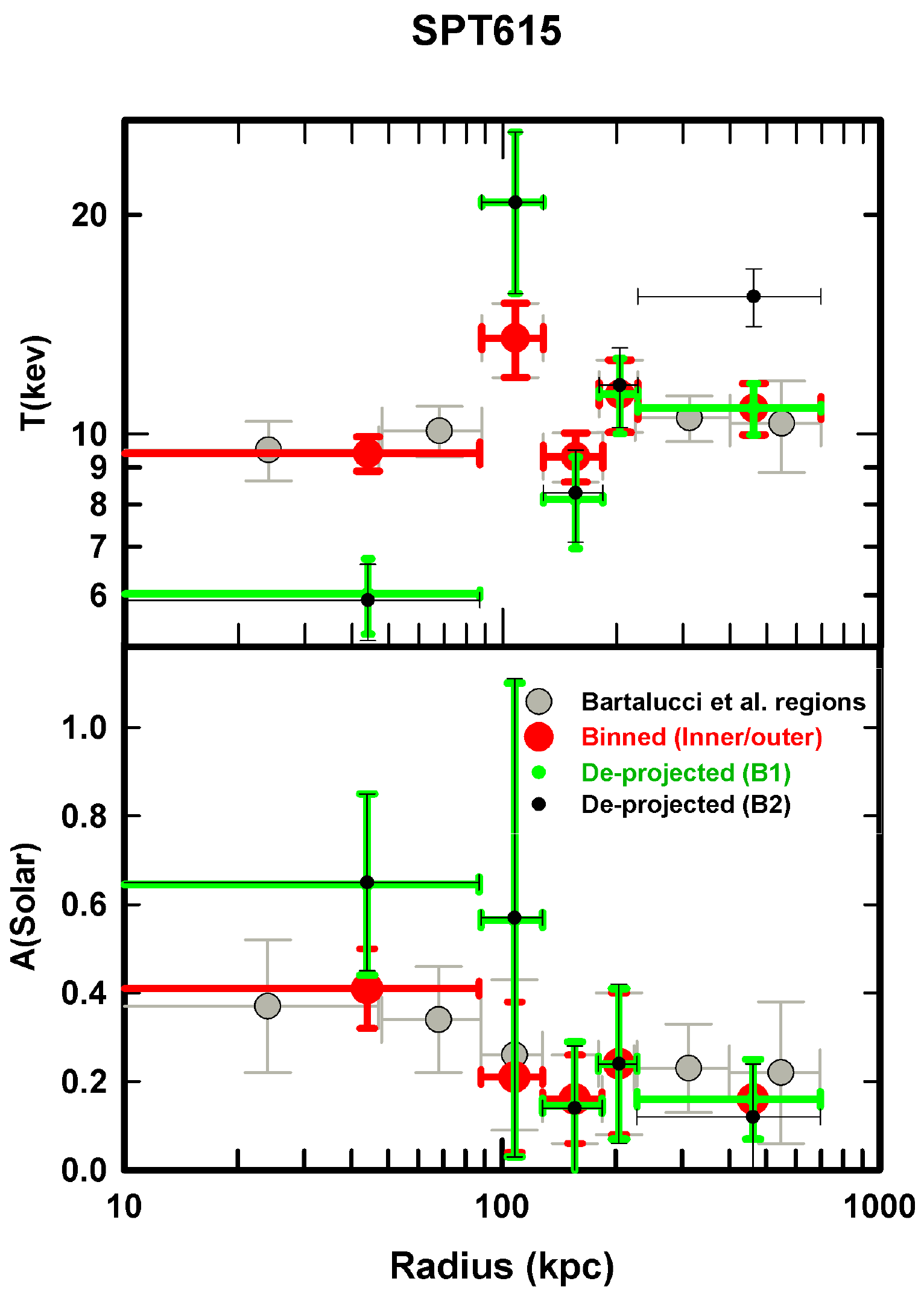}}
\caption{Top: Chandra ACIS-S3 image of SPT0615 with the concentric annuli (white) regions used in \cite{bartalucci2017} and reanalyzed in this work. Also shown two different background regions (B1 and B2) in this  analysis (yellow) to illustrate the possible contamination of the cluster emission at different distances from the center. Bottom: Projected radial profile of temperatures and metal abundances corresponding to the concentric annuli shown in the figure at the top are shown in gray. We also show the same distribution where we join the inner two bins and the outer two bins in red. In green and black, we show the de-projected profiles using the B1 and B2 background regions respectively. North is up. East is left.}
\label{radial5}

\end{figure}

\begin{figure}
\centering
\includegraphics[width=8cm]{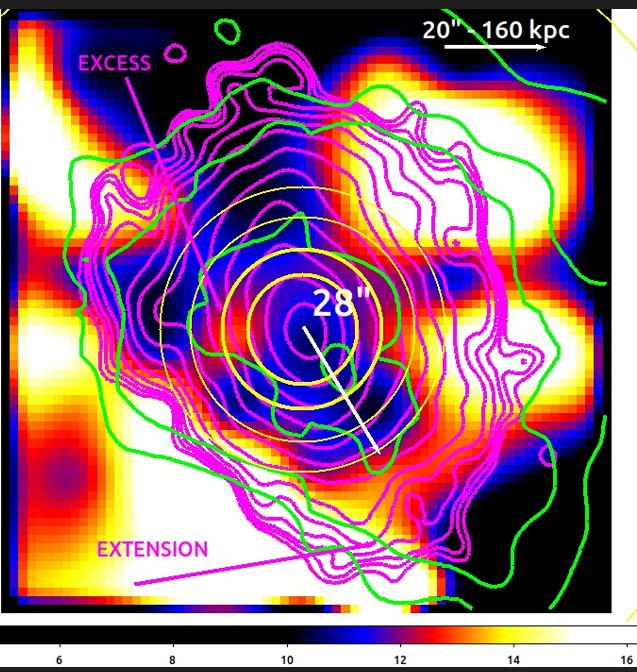}
\includegraphics[width=8cm]{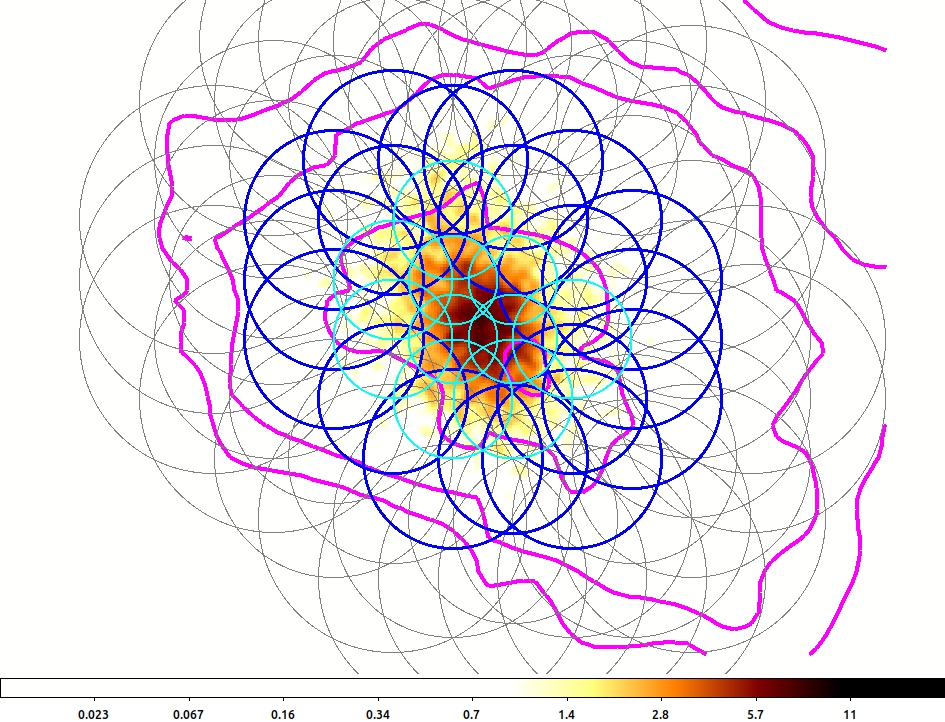}

\caption{
Top: Adaptive smoothed image of the projected intracluster gas temperature in keV.  Surface brightness contours in  {magenta} are the same as in Fig. \ref{x-cont} bottom. Concentric annuli used for radial spectral extraction in Fig. \ref{radial5} are shown in yellow and the annulus corresponding to the temperature spike is in bold. The green contours show the fractional error of the temperatures corresponding to 10\%, 15\%, 25\% and 35\% from inside outwards. North is up. East is left. { In white it is indicated the position of a candidate cold front from a previous merger (Sect. \ref{ximage})}. Bottom: Smoothing overlap characteristic size of regions with radii below 25$\arcsec$, prior to Kriging interpolation. The magenta contours correspond to the same temperature fractional errors plotted in green in the top panel.}
\label{smooth}
\end{figure}

 We extracted the radial distribution of the intracluster gas temperatures (T$_X$) and metal abundances (A$_X$), using circular annuli with the same (or as close as possible) binnage as that used by \cite{bartalucci2017} using local backgrounds (Fig. \ref{radial5} top). We show the results in Fig. \ref{radial5} bottom, where the projected T$_X$ and A$_X$ are seen in gray, and in red we show the same but binning the 2 outermost annuli and the two innermost annuli, since they had similar projected temperatures, to improve statistics (see Table \ref{bestfit_proj}). One can see that there are significant departures from isothermality within 100 kpc--200 kpc from the cluster’s center. To reduce the contamination from the external layers we carried out a standard deprojection, following a simple onion-peel-like procedure with fixed outer layer parameters and normalizations corrected by region area. The results are shown in Table \ref{bestfit_depro} and plotted in Fig. \ref{radial5} bottom. 
 The deprojection increased the significance of the temperature gradient in the 100--200 kpc region and also showed a more clear presence of  {a cooler central region}\footnote{ {We are not convinced that this cooler region constitutes a classic cool core, especially given the temperature distribution shown in Fig. \ref{smooth}}} (Fig. \ref{radial5} bottom in green and black for the two background choices). The region 88--128 kpc (``hot ring'') exhibits extremely high temperatures (T$_X >$15 keV), independent of the background choice, with characteristic values corresponding to  shock regions in very strong mergers such as the Bullet cluster \citep[][]{bullet}, the Pandora cluster \citep[][]{pandora} or Abell 754 \citep[][]{a754}, and others. It should be noted that the temperature of the hot ring as taken from deprojection is likely to be overestimated  {because of the assumption of isotropy for the radial temperature gradient around it. 
 This could also 
 make the cooler central region even cooler as well.} That is a limitation of deprojection technique performance for clusters that have significant amounts of substructures.
 \\
 
Although the overall results are somewhat consistent with those of \cite{bartalucci2017} there is a significant disagreement near the observed hot ring, the exact nature of which cannot be immediately assessed by the radial distribution. However, the presence of such significant temperature substructure, strongly suggests that the system is in the process of, or has passed recently through, merging and that a  purely radial analysis may not be ideal to provide information about the current dynamical state of the cluster. 
 Given that there are, as described above, indications of substructures throughout the cluster, we produced an adaptively smoothed temperature map by performing spectral fittings of circular regions with a minimum of 3000 counts per region, found to be the minimum value, within which we could obtain good enough precision to have discriminatory power, given the cluster's high temperatures. The subsequent gridding method used is based on an interpolation method that calculates a new value for each cell in the regular output matrix for the parameter of interest (T$_X$), from the values of the points in the adjoining cells that are included within a given search radius. The interpolation used follows the Kriging method and has been shown to enhance the mapping of features of the parameters of interest in the intracluster medium successfully  \citep[e.g.,][]{dupkea576,dupkea496}. For the sake of illustration, we show in Fig. \ref{smooth} bottom the smoothing overlap characteristic size of regions with radii below 25$\arcsec$. \\ 

\begin{table}
\centering
\begin{tabular}{p{0.8cm}p{1.1cm}p{1.5cm}p{1.5cm}p{1.2cm} }
 \multicolumn{5}{c}{Best-fit parameters - projected} \\
 \hline
 Region & R$\pm\delta R$&T$_X$ &Abundance & $\chi^2/dof $\\
\hline
  \#&(kpc)& (keV)&(solar)& \\
 \hline
1 & 24$\pm$24 & 9.51$\pm$0.89 & 0.37$\pm$0.15 & 433/434\\
2 & 68$\pm$20 & 10.01$\pm$0.81 & 0.34$\pm$0.15 & 511/602\\
1\&2 & 44$\pm$44 & 9.4$\pm$0.51 & 0.41$\pm$0.09 & 851/954\\
3 & 108$\pm$20 & 13.54$\pm$1.58 & 0.26$\pm$0.17 & 508/550\\
4 & 156$\pm$28 & 9.31$\pm$0.72 & 0.16$\pm$0.10 & 550/612\\
5 & 204$\pm$24 & 11.34$\pm$1.29 & 0.24$\pm$0.16 & 397/460\\
6 & 312$\pm$88 & 10.52$\pm$0.76 &0.23$\pm$0.10 & 899/1049\\
7 & 548$\pm$148 & 10.34$\pm$1.5 & 0.4$\pm$0.12& 936/1092\\
6\&7 & 462$\pm$234 & 10.85$\pm$0.88 & 0.22$\pm$0.16 & 1657/1843\\
\hline
\end{tabular} 
  \caption{Best-fit parameters for the projected spectral simultaneous fittings for all 12 observations.}\label{bestfit_proj}
  \end{table}

\begin{table}
\centering
\begin{tabular}{p{0.8cm}p{1.1cm}p{1.5cm}p{1.5cm}p{1.2cm} }
 \multicolumn{5}{c}{Best-fit parameters - Deprojected} \\
  \hline
 Region & R$\pm\delta R$&T$_X$ &Abundance & $\chi^2_{red} $\\
\hline
  \#&(kpc)& (keV)&(solar)& \\
 \hline
1\&2 & 44$\pm$44 & 6.03$\pm$0.72 & 0.65$\pm$0.21 & 854/954\\
3 & 108$\pm$20 & 20.8$\pm$5.2 & 0.57$\pm$0.54 & 507/550\\
4 & 156$\pm$28 & 8.13$\pm$1.17 & 0.15$\pm$0.15 & 549/612\\
5 & 204$\pm$24 & 11.35$\pm$1.35 & 0.24$\pm$0.17 & 399/460\\
6\&7$^{B1}$ & 462$\pm$234 & 10.85$\pm$0.88 & 0.16$\pm$0.09 & 1657/1843\\
6\&7$^{B2}$ & 462$\pm$234 & 15.5$\pm$1.4 & 0.12$\pm$0.12 & 1803/1843\\
\hline
\end{tabular}
  \caption{Best-fit parameters for the deprojected spectral simultaneous fittings for all 12 observations. The changes with respect to background regions B1 and B2 are seen only for the last anullus and are notated by the indexes B1 and B2 referring to Fig. \ref{radial5}.} \label{bestfit_depro}
  \end{table}

\begin{figure}
\centering
\includegraphics[width=8cm]{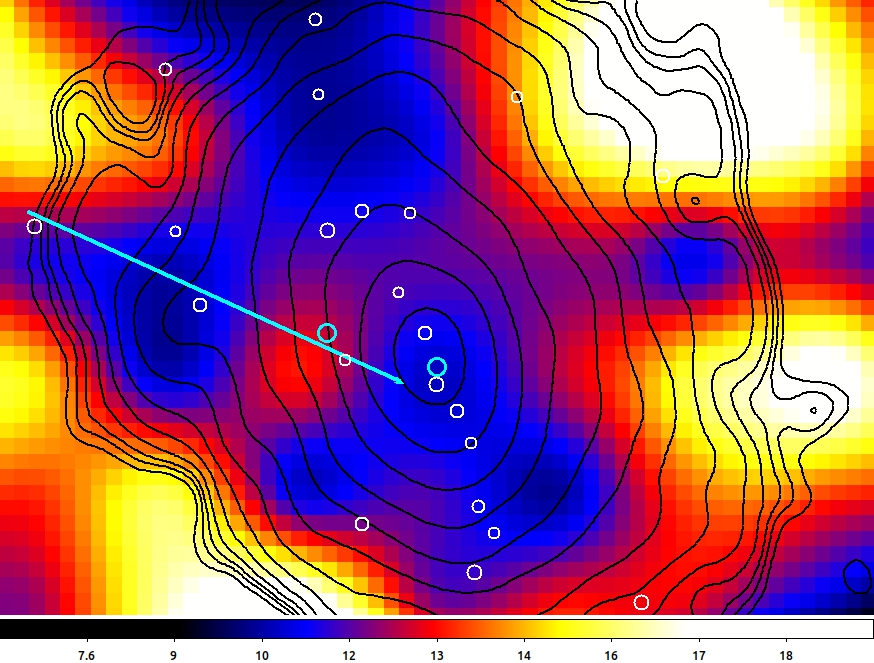}

\caption{Zoom-in the temperature map shown in Fig. \ref{smooth}, with X-ray contours in black and a possible projected merging configuration. The position of the BCG and the large ELG of the hypothesized  incoming cluster are shown in cyan (bold) as well as the other confirmed member galaxies in the region. North is up. East is left.}
\label{dumfuc2}

\end{figure}

 The resulting temperature map is shown in Fig. \ref{smooth} top. The contours shown in green represent different levels of significance, based on the fractional temperature error. Outside the  25\% -- 35\% contours, the fractional errors grow very fast, so that the values are not informative, and can be ignored. However, within the 25\% confidence there is significant presence of substructures. Most notably, a hot ($T_X\sim$13 keV) clump east of the main central BCG, hereafter called "hot clump", which is encompassed by the ``hot ring" annulus, discussed previously. The hot clump is very near the Eastern excess emission, though slightly closer to cluster's core. The actual Eastern excess emission seems to be in a region of  significantly lower temperatures ($T_X < $10 kev).
 The hot clump is also very near the large elliptical luminous galaxy (ELG) with $\Delta_z=0.004$ from the BCG (see Fig. \ref{dumfuc2}). The overall configuration is consistent with a current incoming merging cluster with some significant line-of-sight component, the main central galaxy of which being the Eastern ELG, where the gas is being shock heated to the high temperatures observed in the hot clump. The Eastern excess in this scenario could be attributed to the intracluster medium of the incoming cluster, possibly stripped or not shock heated yet by the merger. We show the suggested configuration in Fig. \ref{dumfuc2}. 
 {There is also a clear anisotropy in the direction of NE--SW at larger scales as well. Particularly interesting, there is a hot shell region in the direction of the southern extension, within the contours of high significance.  
 This substructure, as we hypothesize in the next section, seems to be coincident with a sudden drop in density, characteristic of cold fronts suggesting that a previous recent merger may have happened in the direction NE-SW.}

\subsection{Image}\label{ximage}

\begin{figure}
\centering
\includegraphics[width=9cm]{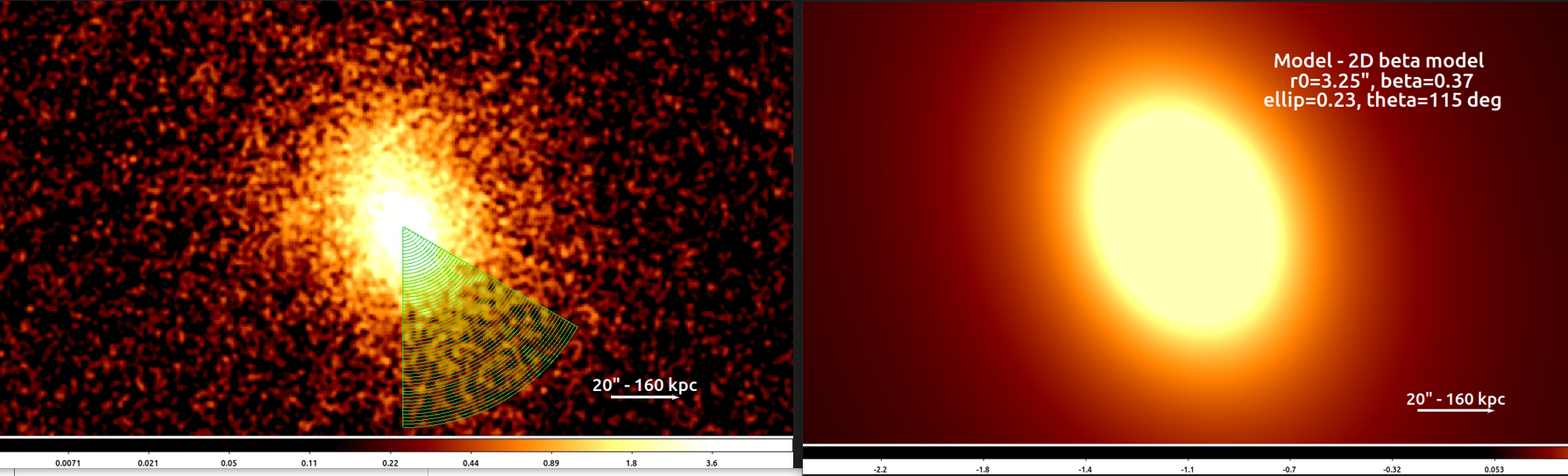}
\includegraphics[width=9cm]{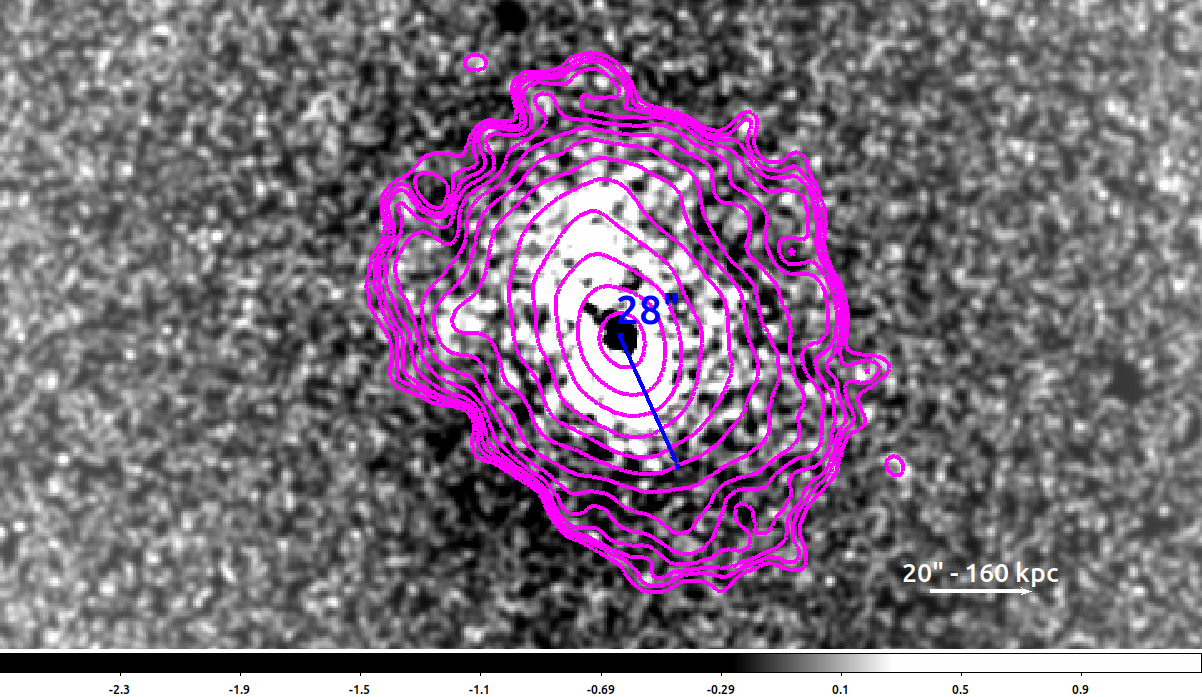}
\caption{ {Top left: Chandra image of SPT0615 used for  2D-fit with a beta model. We also indicate the angular region used to measure the surface brightness profile in Fig. \ref{ccf}. Top right: the beta model and best fit parameters. Bottom: Residual from the fit. We can see the region near the hot core very clearly detached from the other excess over a beta model fit region, which includes a significant part of the central elongation extending 280 kpc to each side from the center.  North is up, East is left.}
}
\label{2dfit}
\end{figure}

 {Even though the exposure time available for this cluster is too low for a detailed photo-spectroscopic analysis of the merger details, it can provide some insights about the recent evolutionary history of this cluster. For example, one can check the level of departure from relaxation by looking at the residual from a simple 2-dimensional beta model image fitting. We created a model with a ellipticity of 0.23 given the strong elongation seen in the X-ray image and a position angle of 115$^{\circ}$. The image, model and residuals from the best fit are shown in Fig. \ref{2dfit}. In Fig. \ref{2dfit} bottom, we can see two interesting features: the confirmation of the eastern excess and the excess emission along the elongation axis. We do not try more complex 2D distributions given the current photon count available, but this excess could be associated with a previous near plane-of-the-sky merger along that elongation axis, where the incoming system's ICM had been stripped/deposited along that path.} \\

\begin{figure}
\centering
\includegraphics[width=8cm]{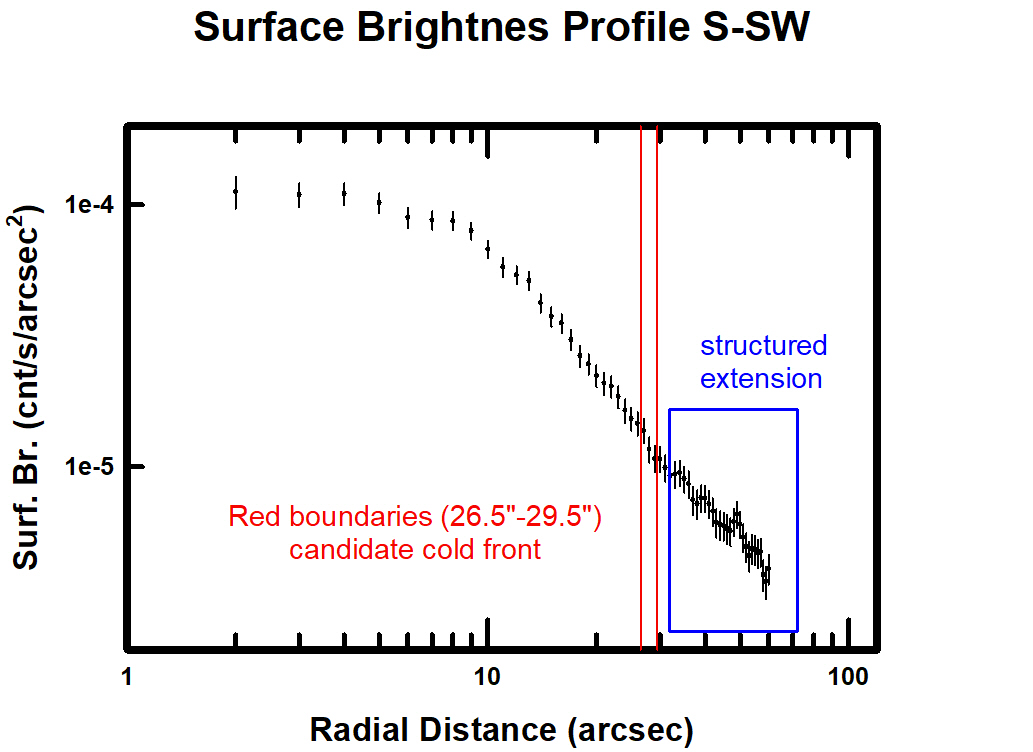}

\caption{ {
Surface brightness profile throughout the angular region shown in Fig. \ref{2dfit} left in the direction of the southern hot shell seen in the temperature map in Fig. \ref{smooth}. A marginal sudden surface brightness decrease is seen about 28\arcsec~ from the cluster's center, coinciding with the inner hot shell. The extended emission past that point is also noted.} }
\label{ccf}
\end{figure}

 {Given the apparent extended emission towards the SW, we extracted the background subtracted surface brightness profile in an angular region towards that direction (Fig. \ref{2dfit} top left) and show the results in Fig. \ref{ccf}. The surface brightness distribution shows a marginally significant drop at about 28\arcsec~ from the center, and some extended excess emission emission with apparent substructures past 40\arcsec. The drop in surface brightness coincides with a raise in temperature of the gas by about 20\% as shown in Figure \ref{smooth}. These characteristics, if confirmed, are consistent with the presence of a cold front \citep{coldfront} due to a previous merger along the elongation axis of the cluster and could be associated to the excess emission in that direction.} 

\section{Discussion}\label{sect:discussion}

\begin{figure*}
\centering
\includegraphics[width=0.9\textwidth]{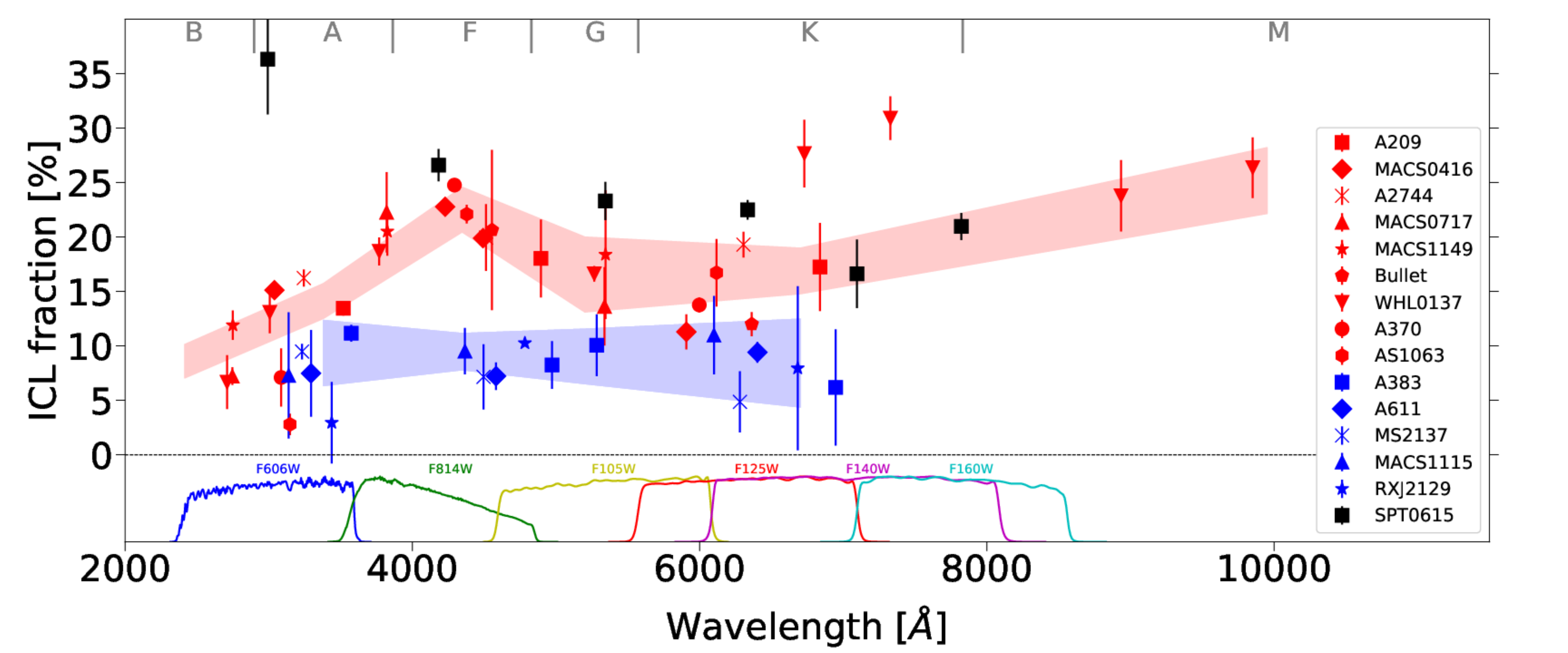}
\caption{ICL fractions of SPT0615 (black squares), as derived by CICLE, compared with a subsample of relaxed (blue) and merging (red) clusters, at intermediate redshift. The plot is divided in six regions according to the peak of emission of the different stellar types, indicated at the top with gray letters.  Red and blue regions indicate the error-weighted average of the measurements within each stellar region, for the merging and relaxed systems, respectively. At the bottom, we plot the transmission curves of the filters used for the analysis of SPT0615. All points are in the rest-frame. }
\label{fig:ICLfractions}
\end{figure*}

In \citet{jimenez-teja2018}, we measured the ICL fractions in three different HST optical filters for a sample of massive clusters that had their dynamical stages well defined by different indicators (X-ray morphology, velocity distribution, presence of tidal features and radio structures, and others). In \citet{jimenez-teja2019,jimenez-teja2021,deoliveira2022}, we expanded the sample adding a few more clusters and confirmed that the  {distribution of ICL fractions measured at different rest-frame optical wavelengths} differs from active (merging) to passive (relaxed) systems. Clusters in this sample spanned the redshift interval $0.02<z<0.56$, which is significantly lower than that the redshift of SPT0615, $z=0.97$. However, as the two-phase scenario of ICL formation is widely accepted (i.e., the ICL is intimately linked to the BCG build-up at $z>1$, while other mechanisms play a major role for $z<1$) one would expect a similar distribution for the ICL fractions of SPT0615 and the intermediate-redshift clusters. 
For the sake of comparison, we plotted in Fig. \ref{fig:ICLfractions} the ICL fractions of SPT0615 (black) and those of the merging (red) and relaxed (blue) clusters of the previous samples. Relaxed systems have nearly constant and low fractions in the rest-frame optical wavelengths, which indicates that the primary sources of ICL are steady (for instance, tidal stripping of member galaxies as they orbit towards the center of the gravitational potential of the cluster, enhanced by dynamical friction). Moreover, constant fractions point to a similar stellar composition of the ICL and the cluster galaxies. Contrarily, merging systems display higher ICL fractions and a distinctive peak between 3800 and 4800 \AA~, which is explained by an excess of A- to F-type stars (as compared with the stellar composition of the galaxies), thrown into the ICL in a short time in a violent regime. The probable channels of these ICL stars are related to the merging stage of the cluster (e.g. tidal stripping of infalling galaxies, preprocessing in groups that are being accreted, etc.). { We  show here that}, 
SPT0615 ICL fractions do not show a trend consistent with a passive cluster (see Fig. \ref{fig:ICLfractions}).\\

 {There are, however, 
many previous works in the literature claiming its relaxed state \citep{planck2011,bartalucci2017,morandi2015,yuan2020}. }For example, \cite{planck2011} confirmed the detection of cluster SPT0615 using a shallow XMM-Newton image, and classified it as relaxed based on its density profile, the nearly symmetric distribution of its hot gas and the offset between the BCG and the X-ray peak. \cite{bartalucci2017} previous Chandra analyses of SPT0615 also argued that the cluster is relaxed, based mostly on an inferred smooth radial temperature profile and the presence of a cool core. However, their measurements of the  centroid shift $\langle\omega\rangle$, defined as the standard deviation of the projected distance between the X-ray peak and the centroid, yield contradictory results depending on the resolution of the X-ray data, i.e., Chandra versus XMM-Newton. Furthermore, as we illustrated above,  $\langle\omega\rangle$ defined from an average may not be representative of the real shift given the structure near the core. Any of the shifts from X-ray peaks to BCG described here would put the cluster out of the relaxed category. Arguments based on flux concentration for this system with XMM are too uncertain to estimate the dynamical state of the cluster \citep[][]{bartalucci2019}. Both \cite{morandi2015} and \cite{bartalucci2017} found a cool core using the Chandra deeper data, while \cite{bulbul2019} using the XMM-Newton observation measured a \textit{higher} temperature for SPT0615 when the core was included than that found by excising it, which is consistent with a lack of cool core.  {Interestingly, it should be noticed that the core-excised region analyzed by \cite{bulbul2019} is large enough to exclude the hot clump found here, so that their results may be compatible with ours}. Additionally, \cite{bartalucci2017} found complex substructures in the cluster core with the Chandra data (not observed with XMM-Newton), including a small low-surface-brightness emission west to the BCG and the X-ray peaks and a horseshoe-like high surface brightness emission surrounding both peaks. More recently, \cite{yuan2020} measured several morphological parameters using Chandra, finding  {inconclusive} conclusions about SPT0615 dynamical stage. With a radically different approach, \cite{connor2019} performed a kinematic analysis of SPT0615 with spectroscopic redshifts, finding evidence for a non-relaxed dynamical state. They found indications for a non-Gaussian velocity dispersion of SPT0615 galaxies which increases with clustercentric radius along with an offset between the BCG and the rest of the cluster's velocity dispersion.\\

Indeed, our measured ICL fraction values for SPT0615 are more consistent with those of merging clusters, although slightly higher than expected in the optical and, an abnormally high in the bluest band (see Fig. \ref{fig:ICLfractions}). At this point, we consider three hypotheses: 1) a  {large contamination of the measured ICL fractions by bright stars, outer stellar haloes of bright galaxies, and/or an incorrect estimation of the background}, 2) a misclassification of this traditionally-believed relaxed cluster, and/or 3) a possible second structure in the line of sight, as hinted by \citet{paterno-mahler2018}. We rule out the first hypothesis since we thoroughly masked out all the pixels that could contain flux from the nearby, bright stars,  {we fitted accurately the galaxies with CHEFs and inspected the residuals after removing them}, and we carefully estimated and removed the background using NoiseChisel (see Sect. \ref{sect_CICLEdata}), which has been proved to provide excellent results in this matter \citep{borlaff2019,haigh2021,kelvin2023}.  {As described in Sect. \ref{sect_ICLfanderror}, the impact of these sources of contamination is included in the error budget, so it is not large enough to explain the excess found in the ICL fractions.} \\

Our detailed X-ray analysis described in Sect. \ref{sect_Xrayresults},  {strongly favours} the non-relaxed scenario for SPT0615. This re-analysis of the Chandra data shows that the cluster is in the process of merging, possibly with multiple components. The most notable being related to the hot clump about 150 kpc east of the center, with temperatures characteristic of shocks seen in violent mergers, and near an ELG, with a radial velocity difference of $\sim$600 km/s with respect to the BCG of the system. Given the temperature distribution and the galaxy distribution to the NE direction, it is suggested that this is a pre-core-crossing merger with a plane of the sky component in the E-S direction as illustrated in Fig. \ref{dumfuc2}. The large scale temperature gradient seen along the main X-ray/galaxies axis is significant, and suggest a secondary merging event,  {along the elongation axis of the cluster, 
although }the current observations are still too shallow to determine a more precise spatial configuration. \\

\begin{figure*}
\centering
\includegraphics[width=\textwidth]{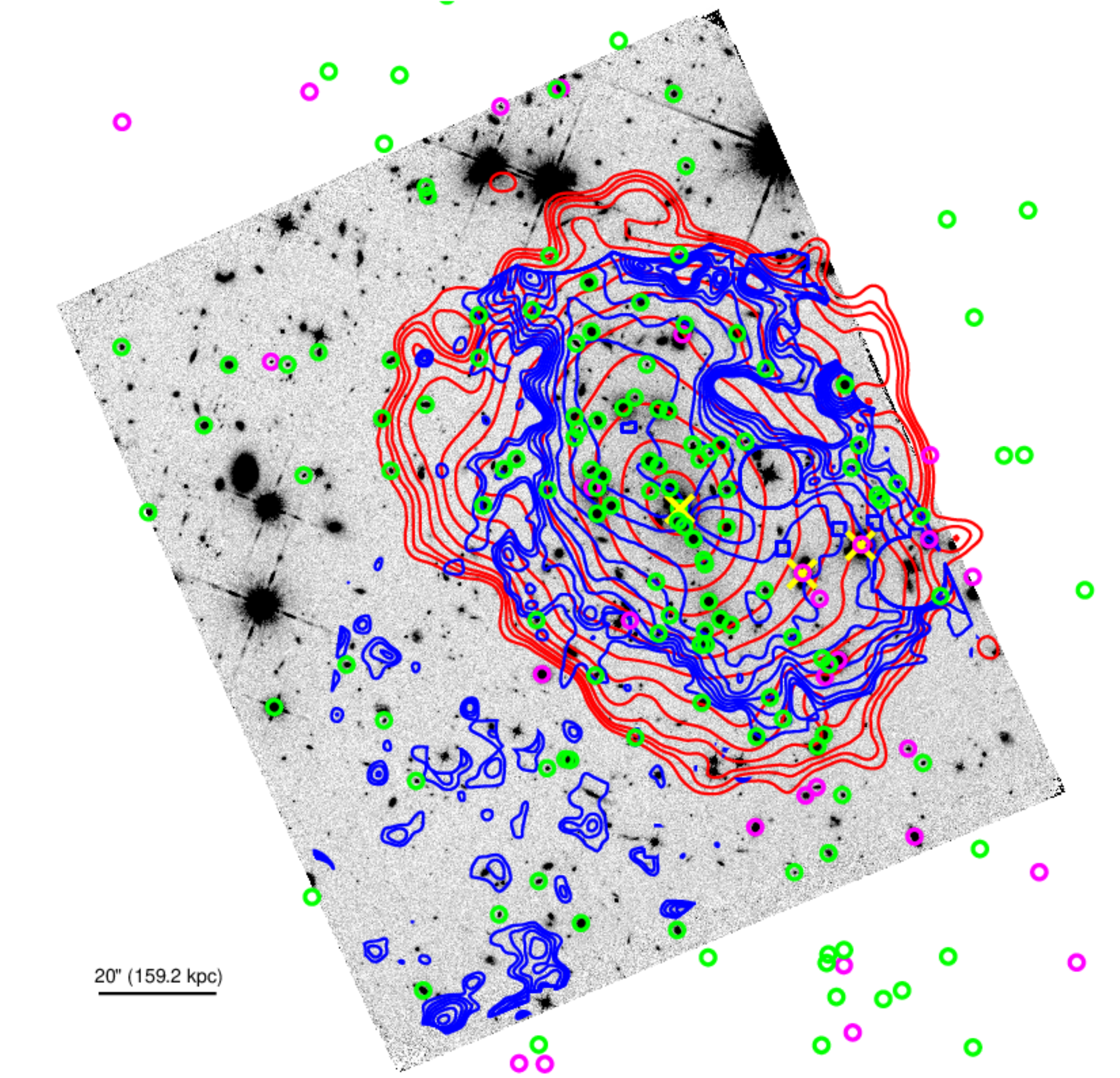}
\caption{Comparison of the ICL, X-ray, and double membership analyses. The contours of the ICL in the F160W are plotted in blue, while those of the hot gas distribution are in red. Green circles mark the position of the galaxies identified as SPT0615 members, and magenta circles indicate the members of the foreground structure at $z\sim 0.44$ according to our machine learning algorithm. SPT0615 BCG and two of the three brightest galaxies of the foreground structure are marked with additional yellow crosses. The third brightest galaxy at $z\sim 0.44$ is out of the F160W footprint, approximately $\sim 654$ kpc (at $z=0.44$) far from the other two, to the southwest. }
\label{fig:members}
\end{figure*}

To further investigate the third hypothesis, the presence of a group at $z\sim 0.44$ with partial projected area overlap with SPT0615, we plot in Fig. \ref{fig:members}, the cluster members ($z=0.97$) in green, the group members ($z\sim 0.44$) in magenta, the ICL contours in the F160W band in blue and the X-ray contour distribution in red. The  {foreground} group members are distributed all along the region of SPT0615, with a certain preference south to the $z=0.97$ BCG. In fact, several of the brightest  {foreground} group members are located close or in the area where we found an excess in the gas temperature { (southwestern extension }in Fig. \ref{smooth} top). However, as several cluster members are also identified in this location, we cannot effectively disentangle if this hot region is due to the possible presence of a  {foreground} group or to the cluster itself. Furthermore, the galaxy distribution near the southwestern excess region at $z\sim 0.44$ kpc seem very disperse over its encompassed $\sim 230$ kpc, with no sign of concentration towards any putative core. It would also be unlikely that the potentially foreground group would have similar high  {intracluster gas} temperatures as that found in SPT0615. As for the excess found both in the X-ray isocontours and the smoothed gas temperature map, east of the BCG (see Figs. \ref{x-cont} bottom and \ref{smooth} top), the ICL contours seem to have an extended distribution or blob matching the position of the putative incoming cluster. As no foreground group members are identified along this region, which instead encompass three bright cluster member galaxies, we conclude that this hot gas excess is due to dynamical activity  {at the SPT0615 redshift }solely, thus confirming its merging stage.\\

Given the difference in richness,  redshift, and the angular offset between the cluster and the possible foreground group, it is very likely that the ICL fractions are primarily dominated by SPT0615. However, some contamination from overlapping regions from the putative  {foreground} group cannot be ruled out with the current data.  {Our previous works showed the importance of measuring total ICL fractions (that is, total photometry as opposed to aperture photometry), especially to be able to use them as reliable indicators of the dynamical stage of the cluster \citep[e.g.][]{jimenez-teja2018,jimenez-teja2021,joo2023}. As a consequence, ``partia'' ICL fractions or ICL fractions measured in certain regions may not be fully representative of the total ICL fractions. This is particularly true in the case of merging clusters, where the spatial distribution of ICL is often non-symmetric and clumpy, which is the case of SPT0615. Taking all these considerations with caution, we make the exercise of measuring the ICL fractions in the (apparently) less polluted region of the ICL, which is North to the BCG. In this region, only two members of the putative foreground group are identified. Additionally, this region contains the main peak of ICL and, therefore, it should enclose a significant part of its total budget. The ICL fractions measured in this region are: $28.61\pm6.36$\% (F606W), $24.2\pm2.7$\% (F814W), $20.8\pm1.9$\% (F105W), $22.2\pm1.1$\% (F125W), $18.4\pm3.4$\% (F140W), and $18.9\pm 1.3$\% (F160W). These partial ICL fractions show: 1) a similar trend to that of total ICL fractions (Table \ref{table:SBlimits_ICLfractions} and Fig. \ref{fig:ICLfractions}), thus proving that SPT0615 is indeed dominating the global values, 2) even more consistency with a merging state (the slight excess in the rest-frame optical F814W and F105W ICL fractions disappears), 3) still an abnormally high ICL fraction in the F606W band.}\\

 {This test suggests that the unusual high value found in the rest-frame in near UV may indeed have a physical origin. We can speculate that this component} might come from stripped, early-type 
stars that were formed very recently during the currently ongoing  
merger. 
An upper age limit for the current merger can potentially be set by the half-life of the stars that mainly contribute in the near UV. If the UV excess 
is primarily formed by A0-type main-sequence stars, this would imply an upper limit of $\sim$ 0.5-0.6 Gyr, their typical half-life. A0 stars have a typical color $B-V=0$ and this color is equivalent to that seen in late type galaxies (Sd), e.g. outer spiral disks and irregulars. Thus, if not much pre-processing has operated in infalling halos, blue galaxies could be found still entering the cluster potential and injecting these stars into the ICL. Additionally, the star formation-density relation observed in the local Universe is reversed at $z\sim 1$: the fraction of star-forming galaxies in dense environments is significantly higher than in nearby clusters, which are primarily dominated by red, evolved galaxies \citep{elbaz2007}. That would imply that the stars stripped from these star-forming galaxies into the ICL are bluer when compared with those of low- and intermediate-redshift clusters (note that all clusters studied in previous works spanned the interval $0.18<z<0.56$ while SPT0615 is at $z=0.97$, see Fig. \ref{fig:ICLfractions}). 

\section{Conclusions} \label{sect:conclusions}

Despite being a high-redshift cluster, SPT0615, has been  {previously considered} to be relaxed. Here, we revisited this scenario using two completely independent and reliable indicators of the dynamical stage: the ICL fraction and a revised analysis of the hot gas distribution. Previous work showed that the ICL fraction measured at different optical and infrared wavelengths, exhibits a unique marker that differentiated active (merging) from relaxed (passive) systems \citep{jimenez-teja2018,jimenez-teja2019,jimenez-teja2021,deoliveira2022,dupke2022}. This has been verified in many massive clusters at low- and intermediate-redshift ($0.02<z<0.56$)  systems. The merging signature was identified as a characteristic excess in the rest-frame ICL fractions measured roughly in the peak wavelength correspondent to stars of A to F spectral types, i.e., between 3800 and 4800 \AA. For SPT0615, we found an ICL fraction distribution that does not follow that typical of the relaxed clusters, but it is more consistent with an active system. Actually, even showing indications of an excess in the UV ICL fraction, which cannot be explained with a merger solely, requiring the radial gradient of average spectral type observed in nearby galaxies to include very early type stars at large radii. \\

We also performed an independent analysis of the existing X-ray data, and noticed that there were significant departures from the expected temperature distribution for relaxed CC clusters, with near-core intracluster gas temperatures corresponding to those of very strong mergers. Subsequently, we produced a full temperature map, which allowed us to detect several substructures (at least two major ones), that are consistent with the presence of multiple mergers, one currently in a pre-core crossing stage, which created shock heated hot clump between the BCGs, and the other likely to be remnant of a previous merger. \\

As the photometric redshift distribution of the galaxies in the region of SPT0615 showed two main peaks, one at $z\sim 0.97$ corresponding to the cluster and another one at $z\sim 0.44$, we performed a double cluster membership analysis using a machine learning algorithm. The results show two clear red sequences with 176 and 36 members, respectively. The projected distribution of the cluster and  {foreground} group galaxies partially overlap, with a larger concentration of group members southwest to the cluster BCG.  {However, the ICL analysis performed in regions with little or no projected overlap does not indicate significant change in the overall results. Furthermore, the lack of concentration around some putative foreground BCG, by either galaxies or intracluster gas and the general spread of the foreground galaxies suggest that there is no significant contamination.} \\

We list here the main results found from the HST and X-ray analyses:

\begin{itemize}
    \item the ICL contours show an extended ICL, elongated along an axis with a position angle of $\sim 30$ degrees, and two main clumps north and southwest of the BCG.
    \item ICL fractions range from 16.6 to 36.3\%, with a spectral distribution consistent with that of merging clusters, although slightly higher than expected in the rest-frame optical bands. 
    \item ICL fraction in the bluest filter, F606W, is abnormally high when it is compared with that of low- and intermediate-redshift clusters, either merging or relaxed, but consistent with an extension of the ICL fraction peak seen in merging clusters. This would be consistent with observing a cluster merger with a upper limit age of $\sim$ 0.5 Gyr if the injected ICL stars are of very early type, A0 or earlier. This would imply that late-type galaxies did not suffer much pre-processing before infalling into the cluster potential. Also, the reversal of the star formation-density relation in the distant Universe can provoke a higher presence of bluer stars in the ICL.  
    \item The X-ray surface brightness shows an elongation along a position angle of $\sim25$ degrees, same as seen in the optical  {analysis, and along which there is significant brightness excess with respect to a standard beta profile.}
    \item the de-projected temperature profile confirms a cool core but shows a strong temperature fluctuations within 100-200 kpc. 
    \item adaptive smoothed temperature map shows significant azimuthal substructures, in particular a hot clump about 150 kpc far from the cluster center near a ELG, consistent with a strong ongoing (pre core-crossing) merger with temperatures consistent with shock heating gas found in other strong mergers. This clump { traces well 
    an ICL filament} which is associated to three bright cluster member galaxies, which corroborates the ongoing merging hypothesis.
    \item There is significant evidence of other temperature substructures consistent with previous or ongoing secondary mergers.  {In particular, the surface brightness profile towards the southwest along the elongation axis of the cluster seems to have a slight drop at $\sim$ 28\arcsec~  from the center, right before a abrupt temperature rise, which is characteristic of cold fronts of the merging (not sloshing) type \citep{dupkea496,cfrev}. If confirmed by future observations, this would indicate that a previous merger happened recently along the elongation axis and that the gaseous core of the incoming system would be $\sim$ 280 kpc southwest of the cluster's center.}
\end{itemize}

This work highlights the power of the ICL and, in particular, the ICL fractions, to determine the dynamical stage of clusters. CICLE provides unbiased measurements of the ICL fractions, able to discover anomalies that can  be associated to merging activity and/or projection effects, 
in particular when coupled with X-ray observations  {with high spatial resolution such as that of Chandra. This is particularly important to study the evolution of high-z clusters, given their small angular sizes. The forthcoming JWST observations\footnote{ {https://www.stsci.edu/jwst/science-execution/program-information.html?id=4212}} of SPT0615 is poised to provide crucial insights into determining the merging configuration.}
 \\

\section*{Acknowledgements}

 {All authors sincerely thank the anonymous referee for his/her useful and kind comments, which have certainly improved the quality of this manuscript.}
Y.J-T. acknowledges financial support from the European Union’s Horizon 2020 research and innovation programme under the Marie Skłodowska-Curie grant agreement No 898633 and the MSCA IF Extensions Program of the Spanish National Research Council (CSIC). R.A.D. acknowledges partial support from the CNPq grants 308105/2018-4 \& 312565/2022-4. J.M.V. thanks support from project PID2019- 107408GB-C44 (Spanish Ministerio de Ciencia e Innovación). Y.J-T., R.A.D., and J.M.V. acknowledge support from the State Agency for Research of the Spanish MCIU through the
"Center of Excellence Severo Ochoa" award to the Instituto de Astrofísica de Andalucía (SEV-2017-0709) and grant CEX2021-001131-S funded by MCIN/AEI/ 10.13039/501100011033. P.A.A.L. thanks the support of CNPq (grants 433938/2018-8 e 312460/2021-0) and FAPERJ (grant E-26/200.545/2023). We thank Dr. Rebeca Batalha for helpful discussions.
This work is based on observations taken by the RELICS Treasury Program (GO 14096) with the NASA/ESA HST, which is operated by the Association of Universities for Research in Astronomy, Inc., under NASA contract NAS5-26555.

\bibliography{main.bib}{}
\bibliographystyle{aasjournal}

\end{document}